\documentclass[twocolumn,superscriptaddress,showpacs,amsmath,amssymb,pra,longbibliography]{revtex4-1}
\usepackage[pdftex]{graphicx} 
\usepackage{dcolumn}  
\usepackage{bm}       
\usepackage[usenames,dvipsnames]{color}
\definecolor{URLCOL}{rgb}{0,0.52,0.83} 
\definecolor{LINKCOL}{rgb}{0.05,0.5,0} 
\definecolor{orange}{rgb}{0.6,0.3,0} 
\definecolor{CITECOL}{rgb}{0.25,0,0.48} 
\usepackage{epstopdf}
\usepackage{amsmath}
\usepackage[table,xcdraw]{xcolor}
\usepackage{amsfonts}
\usepackage{booktabs}
\usepackage{siunitx}
\usepackage{multirow}
\usepackage[pdftex,bookmarks,breaklinks,bookmarksopen,bookmarksnumbered,colorlinks,linkcolor=LINKCOL,linktocpage,citecolor=CITECOL,urlcolor=URLCOL,pdfpagemode=UseOutline,pdftex]{hyperref}
\usepackage{nicefrac}

\usepackage{fancyhdr}
\fancypagestyle{titlepage}
{\chead{file: \jobname}\rhead{Total pages: \pageref{page:end}}}

\definecolor{SEGreen}{rgb}{0.1,0.7,0.5}

\usepackage{booktabs}
\usepackage{subfigure}
\usepackage{natbib}

\definecolor{TITLECOL}{rgb}{0.1,0.2,0.7} 
\definecolor{SECOL}{rgb}{0.1,0.2,0.7} 
\definecolor{CONTENTSCOL}{rgb}{0.1,0.2,0.7} 
\definecolor{SSECOL}{rgb}{0.25,0,0.48} 
\definecolor{SSSECOL}{rgb}{0.2,0.08,0.53} 
\definecolor{FINCOL}{rgb}{0.01,0.3,0.07} 

\def\coloredtitle#1{\title{\textcolor{TITLECOL}{#1}}} 
\def\coloredauthor#1{\author{\textcolor{CITECOL}{#1}}} 

\definecolor{URLCOL}{rgb}{0,0.17,0.43} 
\definecolor{LINKCOL}{rgb}{0.05,0.4,0} 
\definecolor{CITECOL}{rgb}{0.35,0,0.48} 

\def\bea{\begin{eqnarray}}
\def\eea{\end{eqnarray}}
\def\ben{\begin{equation}}
\def\een{\end{equation}}
\def\benu{\begin{enumerate}}
\def\enu{\end{enumerate}}

\def\bei{\begin{itemize}}
\def\eei{\end{itemize}}
\def\beit{\begin{itemize}}
\def\eit{\end{itemize}}
\def\benu{\begin{enumerate}}
\def\enu{\end{enumerate}}

\def\sec#1{\section{\textcolor{SECOL}{#1}}}

\begin{document}

\coloredtitle{
Electronic structure and optical properties of halide double perovskites from a Wannier-localized optimally-tuned screened range-separated hybrid functional
}
\coloredauthor{Francisca Sagredo}
\affiliation{Materials Sciences Division, Lawrence Berkeley National Laboratory, Berkeley, CA 94720}
\affiliation{Department of Physics, University of California, Berkeley, CA 94720, USA}

\coloredauthor{Stephen E. Gant}
\affiliation{Materials Sciences Division, Lawrence Berkeley National Laboratory, Berkeley, CA 94720}
\affiliation{Department of Physics, University of California, Berkeley, CA 94720, USA}

\coloredauthor{Guy Ohad}
\affiliation{Department of Molecular Chemistry and Materials Science, Weizmann Institute of Science, Rehovoth 76100, Israel}

\coloredauthor{Jonah B. Haber}
\affiliation{Materials Sciences Division, Lawrence Berkeley National Laboratory, Berkeley, CA 94720}
\affiliation{Department of Physics, University of California, Berkeley, CA 94720, USA}

\coloredauthor{ Marina R. Filip}
\affiliation{Department of Physics, University of Oxford, Oxford OX1 3PJ, United Kingdom}

\coloredauthor{Leeor Kronik}
\affiliation{Department of Molecular Chemistry and Materials Science, Weizmann Institute of Science, Rehovoth 76100, Israel}
\coloredauthor{Jeffrey B. Neaton}
\affiliation{Materials Sciences Division, Lawrence Berkeley National Laboratory, Berkeley, CA 94720}
\affiliation{Department of Physics, University of California, Berkeley, CA 94720, USA}
\affiliation{Kavli Energy NanoScience Institute at Berkeley, Berkeley, CA 94720, USA}

\date{\today}

\begin{abstract}
Halide double perovskites are a chemically-diverse and growing class of compound semiconductors that are promising for optoelectronic applications. However, the prediction of their fundamental gaps and optical properties with density functional theory (DFT) and {\it ab initio} many-body perturbation theory has been a significant challenge. Recently, a nonempirical Wannier-localized optimally-tuned screened range-separated hybrid (WOT-SRSH) functional has been shown to accurately produce the fundamental band gaps of a wide set of semiconductors and insulators, including lead halide perovskites. Here we apply the WOT-SRSH functional to five halide double perovskites, and compare the results with those obtained from other known functionals and previous $GW$ calculations. We also use the approach as a starting point for $GW$ calculations and we compute the band structures and optical absorption spectrum for Cs\textsubscript{2}Ag{Bi}Br\textsubscript{6}, using both time-dependent DFT and the $GW$-Bethe-Salpeter equation approach. We show that the WOT-SRSH functional leads to accurate fundamental and optical band gaps, as well as optical absorption spectra, consistent with spectroscopic measurements, thereby establishing WOT-SRSH as a viable method for the accurate prediction of optoelectronic properties of halide double perovskites.
\end{abstract}

\maketitle

\sec{Introduction}

Over the past decade, halide double perovskites have emerged as potential lead-free alternatives to light absorbing lead halide perovskites \cite{MBWW16,VFHS16,WDSX16,VHMS17,SLSB18,FLVA16,FG18,VFHS16,SHLK16,DWSK16,DLCR18,YCYH18}. These semiconducting materials possess the compositional flexibility of the perovskite structure, and can exhibit greater thermodynamic stability compared to other perovskites \cite{FLVA16,FG18}. Lower effective masses \cite{VFHS16}, small gaps in the visible range, and long recombination lifetimes \cite{SHLK16} make these materials highly favorable as potential solar absorbers \cite{DWSK16,DLCR18,YCYH18}. 

\begin{figure}[!h]
\includegraphics[width=.7\linewidth, angle=270]{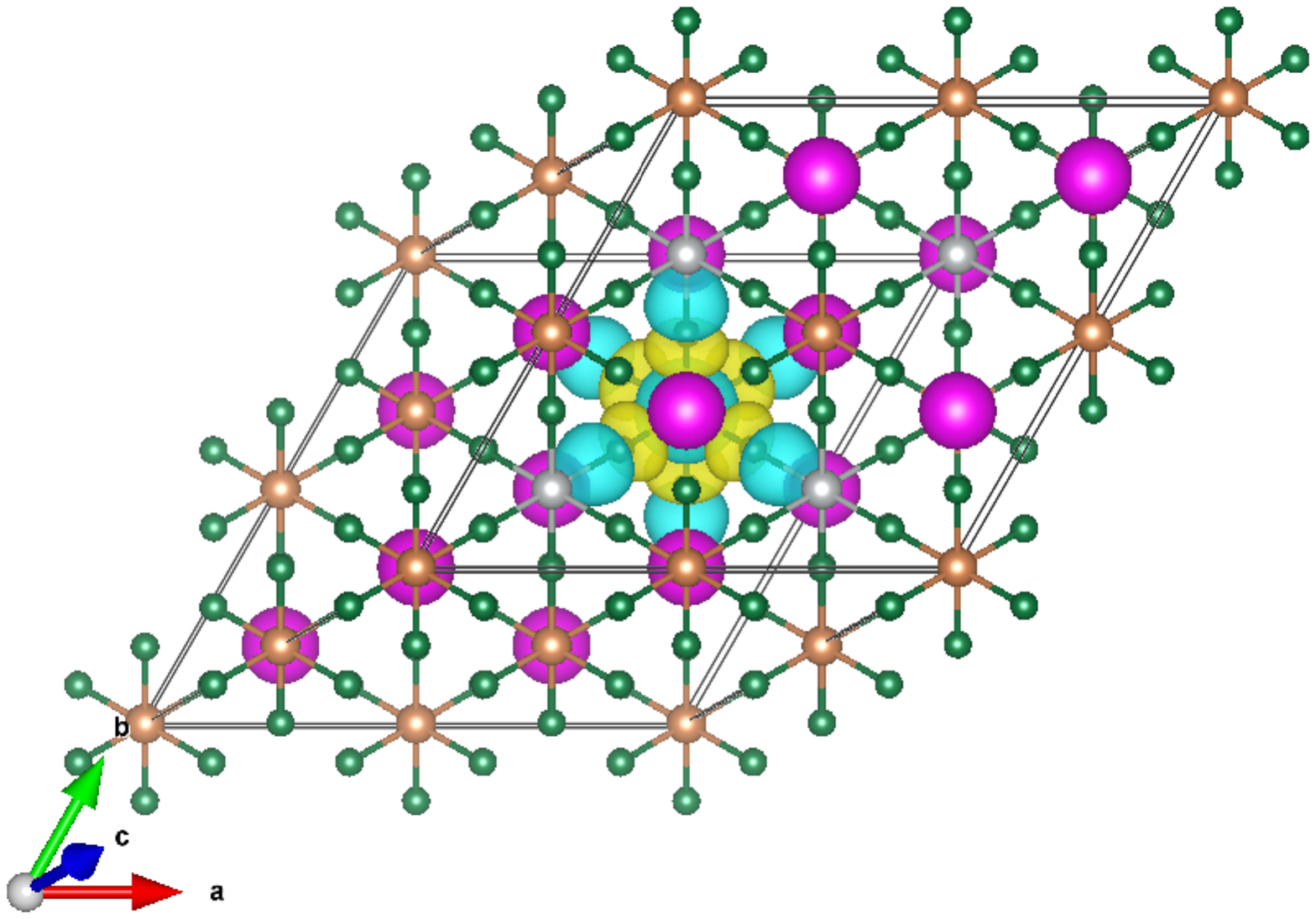}
\caption{Schematic view of the cubic, $F\bar{m}3m$ phase of the halide double perovskite, Cs\textsubscript{2}Ag{Sb}Br\textsubscript{6}. Ag atoms are denoted in silver, and Cs in orange. The Wannier function used for the tuning is seen in blue and yellow; it comprises a localized \textit{s}-orbital on the Sb (magenta) atom, with surrounding \textit{p}-orbitals localized on the Cl (green) atoms, which is in agreement with orbital contributions (\textit{e.g.} orbital makeup) of the valence band maximum predicted in Ref.~\cite{BFLB21}.}
\label{fig wann}
\end{figure}

The optoelectronic properties of halide double perovskites have generated significant theoretical and computational interest, with most calculations of halide double perovskites relying on density functional theory (DFT) \cite{HK64,KS65} and \textit{ab initio} many body perturbation theory (MBPT) within the \textit{GW} approximation and the Bethe-Salpeter equation (BSE) approach \cite{RL00,ARDO98}. These methods are vital tools for the prediction of critical quantities, \textit{e.g.} band gaps, exciton binding energies, effective masses, and optical absorption spectra \cite{H65,HL86} of halide double perovskites \cite{BFLB21,BCFL23}. However, state-of-the art Green's function based approaches are  sensitive to the underlying (generalized) Kohn-Sham ``starting point" used in computationally-efficient ``one-shot" $G_0W_0$ MBPT calculations \cite{BM13,FFBS07,LRN19,VGGR17,RQNF05}. Within Kohn-Sham DFT \cite{KS65}, semilocal and local exchange and correlation (XC) functionals, along with hybrid functionals in the generalized KS scheme \cite{PYBY17}, have been a popular and necessary tool for approximating the fundamental band gaps of material systems, even given their known deficiencies. These XC functionals (including popular global hybrids) are also used to approximate the optical properties of material systems using the time-dependent DFT (TDDFT) framework \cite{RG84,CU12,M16}. But these XC functionals are known to severely underestimate the fundamental gap, mainly due to the missing derivative discontinuity \cite{PPLB82,SAGK12,KSRB12,PL84,SS83}. Thus, the need for an XC functional that is both non-empirical and beyond the semilocal and local level, and that can alleviate some of the deficiencies present in existing XC approximations, is current. 
\begin{figure}
    \centering
    \subfigure(a) {\includegraphics[width=.94\linewidth]{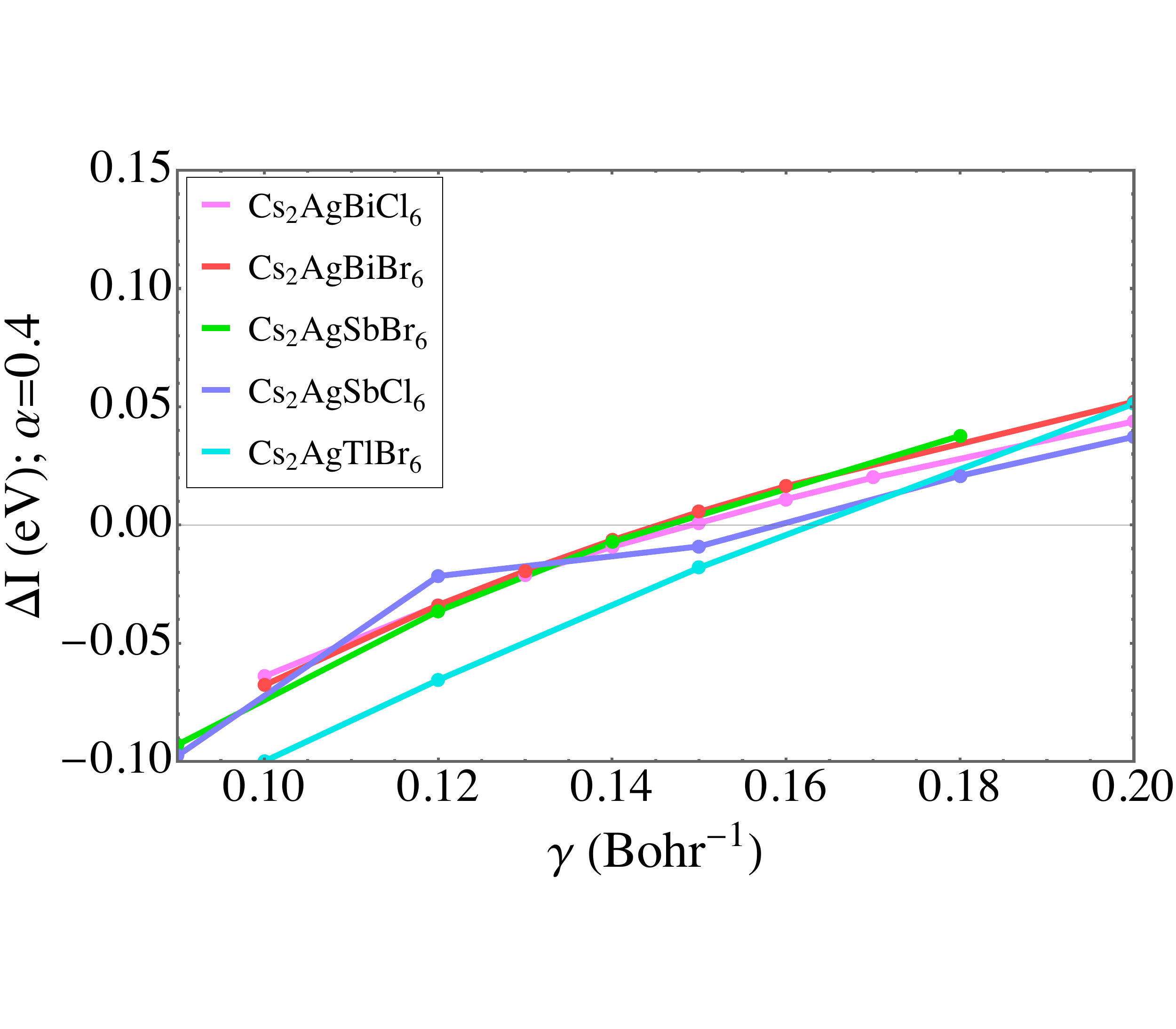}} 
    \subfigure(b) {\includegraphics[width=.89\linewidth]{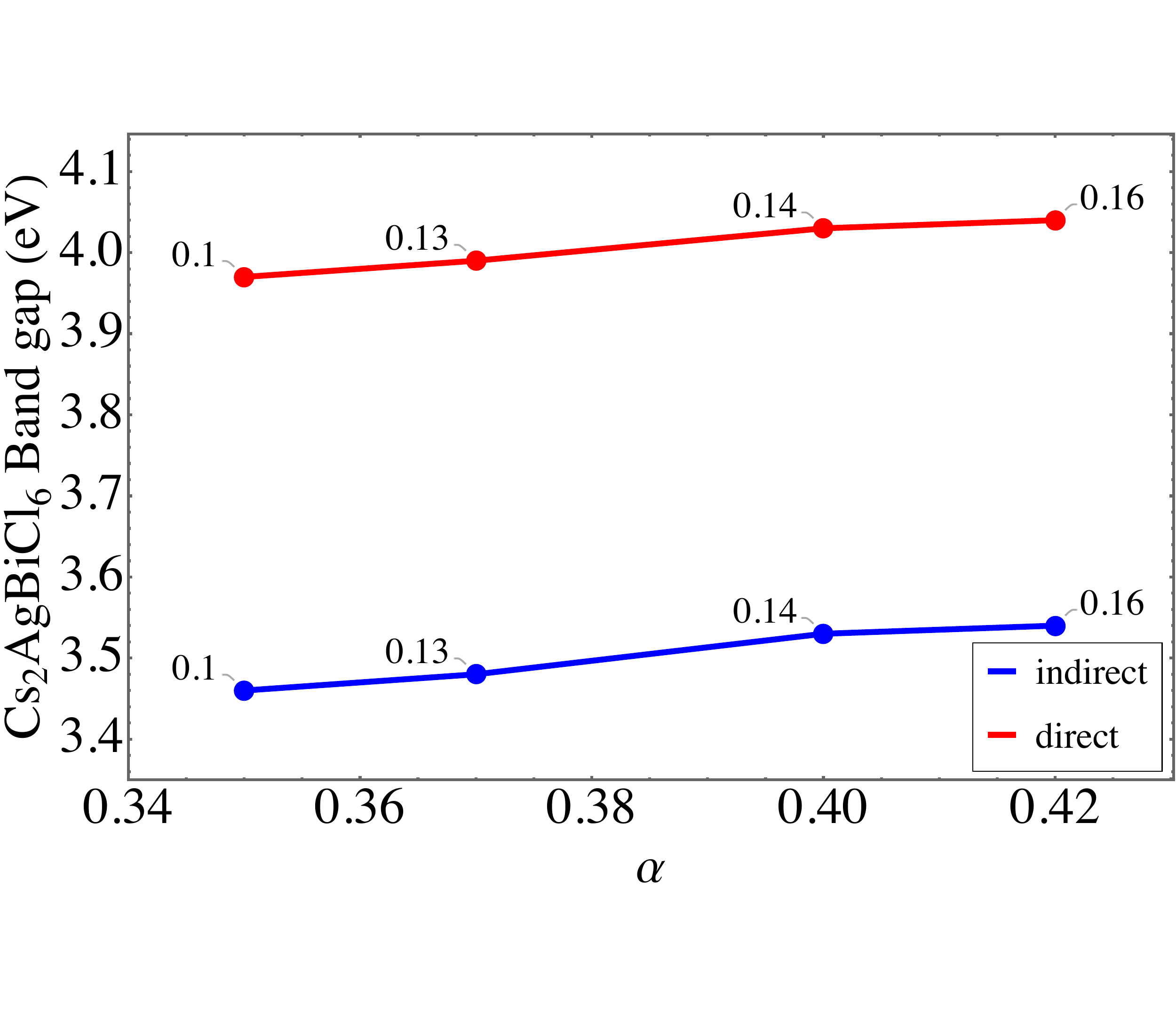}}
    \subfigure(c){\includegraphics[width=.89\linewidth]{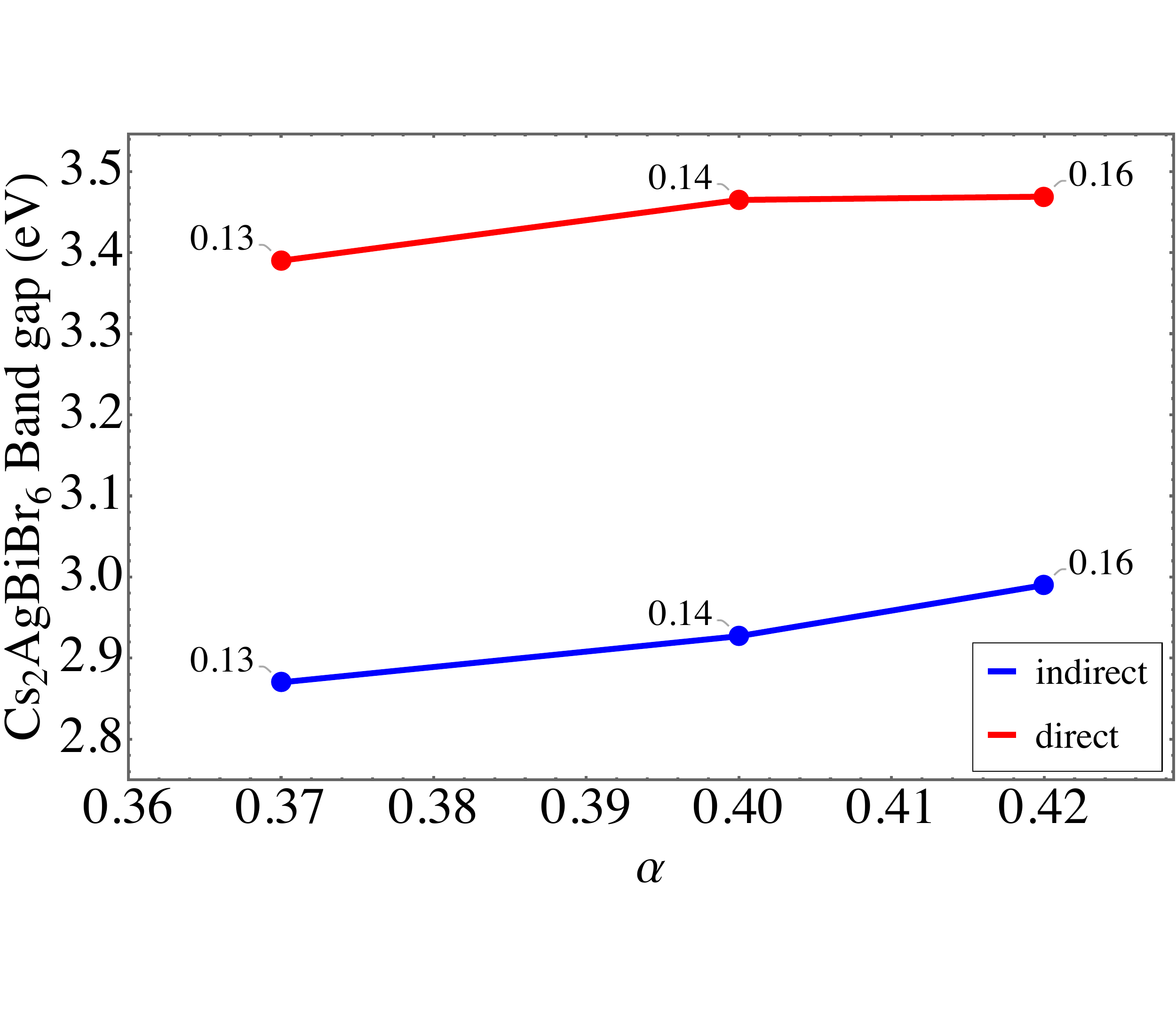}}
    \caption{(a) $\gamma$ tuning plots, with $\alpha =0.4$, for Cs\textsubscript{2}Ag{Bi}Cl\textsubscript{6}, Cs\textsubscript{2}Ag{Bi}Br\textsubscript{6}, Cs\textsubscript{2}Ag{Sb}Cl\textsubscript{6}, Cs\textsubscript{2}Ag{Sb}Br\textsubscript{6}, and Cs\textsubscript{2}Ag{Tl}Br\textsubscript{6}. Variation in band gap (eV) with respect to ($\alpha$, $\gamma$) pairs for (b) Cs\textsubscript{2}Ag{Bi}Br\textsubscript{6}, and (c) Cs\textsubscript{2}Ag{Bi}Cl\textsubscript{6}. Labeled points on the curves correspond to the tuned $\gamma$ ($\text{Bohr}^{-1}$) for the value of $\alpha$. }
    \label{fig tuning}
\end{figure}

To date, there is no single local, semilocal, global hybrid functional, alone or in conjunction with a \textit{GW} approach, that has accurately and consistently predicted fundamental gaps or related properties of halide double perovskites \cite{LRN19}. Their chemical complexity and the presence of heavy elements pose significant challenges \cite{FHHA16}. Furthermore, some of these materials are known to have non-negligible exciton binding energies of $\sim 0.2-0.6$ eV \cite{BFLB21,BCFL23}, complicating the interpretation of experimental data; values extracted from optical absorption spectra that are obtained from fitting procedures with models that do not account for electron-hole interactions can be in error. While significant progress has been made in synthesizing these material systems, experiments report conflicting gaps \cite{VFHS16,FHHA16,SHLH16}, potentially calling for a reanalysis of measured data with models that account for electron-hole interactions \cite{WBSL2021}. An accurate XC functional would provide a deeper understanding of these issues and enable greater predictive power of potential as-yet-unsynthesized or characterized halide double perovskites. 

Recently, a non-empirical range separated hybrid functional was proposed for the prediction of fundamental band gaps and has been benchmarked on  group III-V semi-conductors and insulators \cite{WOHF21}, lead halide perovskites \cite{OWGC22}, and closed-shell metal oxides \cite{OG2023}. This Wannier-localized optimally-tuned screened range separated hybrid (WOT-SRSH) functional has yielded  fundamental band gaps with a mean absolute error (MAE) relative to experiment of $0.08$ eV for a set of typical semiconductors and insulators \cite{WOHF21}, and $\sim 0.1$ eV for lead halide perovskites \cite{OWGC22}, and has even been used as an optimal starting point for MBPT calculations \cite{GHFS22,OG2023}. The parameters associated with this hybrid functional are computed non-empirically on a system-by-system basis to satisfy an ionization potential \textit{ansatz} and to enforce the correct asymptotic screening limit of the Coulomb potential in solids. While use of other advanced functionals can certainly improve the accuracy of band gaps \cite{YU13,ZGG19,CMRP18,TLKP20,DFPL10,DFM14,MCRP18,NCFM18,MWSY22,LCDN23}, the search for a scheme that is accurate to within the experimental uncertainty in a nonempirical fashion for halide double perovskites is ongoing. Here we apply the WOT-SRSH XC functional to compute the fundamental band gaps of five double halide perovskites, namely Cs\textsubscript{2}Ag{Bi}Br\textsubscript{6}, Cs\textsubscript{2}Ag{Bi}Cl\textsubscript{6}, Cs\textsubscript{2}Ag{Sb}Br\textsubscript{6}, Cs\textsubscript{2}Ag{Sb}Cl\textsubscript{6}, and Cs\textsubscript{2}Ag{Tl}Br\textsubscript{6}, and also use it as a starting point for $G_0W_0$ calculations. We compare with common density functionals, previous $GW$ calculations, and recent experimental results, where appropriate. We also use the WOT-SRSH functional to obtain band structures and optical absorption spectra for Cs\textsubscript{2}Ag{Bi}Br\textsubscript{6}, comparing with previously reported optical gaps and with \textit{ab initio} $GW$-BSE calculations. We find that the WOT-SRSH functional predicts fundamental gaps for these systems that are, to date, the most consistent with experiment when accounting for known exciton binding energies and modulo any effects of lattice vibrations.
 
\sec{The WOT-SRSH functional}
\label{sec:WOTSRSH_func}
The screened-range separated hybrid (SRSH) \cite{RSJB13} functional depends on three parameters: the fraction of short-range exact (Fock) exchange, $\alpha$; the range-separation parameter, $\gamma$; and the orientationally averaged high-frequency dielectric constant, $\varepsilon_{\infty}$. These three parameters are introduced into the exchange contribution of the Coulomb interaction via the identity \cite{RSJB13}  
\ben
\begin{split}
\frac{1}{r}= \overbrace{\alpha \frac{\text{erfc}(\gamma r)}{r}}^{XX, \text{SR}} + \overbrace{(1-\alpha) \frac{\text{erfc}(\gamma r)}{r}}^{\text{KS}_{X}, \text{SR}} + \\
\underbrace{\frac{1}{\varepsilon_{\infty}} \frac{ \text{erf}(\gamma r)}{r}}_{XX, \text{LR}} + \underbrace{\bigg(1-\frac{1}{\varepsilon_{\infty}}\bigg)\frac{\text{erf} (\gamma r)}{r}}_{\text{KS}_{X}, \text{LR}} .
\label{eq.1/r} 
\end{split}
\een
In Eq.~(\ref{eq.1/r}), the first two terms correspond to short-range (SR) interactions, and the latter two terms to long range (LR) interactions.  The first and third terms are treated using exact exchange ($\text{xx}$), and the second and fourth terms are treated using Kohn-Sham (KS), semilocal exchange ($\text{KS}_{X}$). Choosing $\varepsilon_{\infty}$ as the prefactor of the LR exact exchange ensures that the many-body interaction tends to the correct physical form in the asymptotic potential \cite{WOHF21,OWGC22,OG2023,GHFS22,RSJB13}.  More generally, the SRSH XC functional using semilocal correlation ($\text{KS}_{C}$) can be written as 

\ben
\begin{split}
E^{\text{SRSH}}_{_{XC}}(\alpha,\gamma,\varepsilon_{\infty})=\alpha E_{_{XX}}^{\text{SR}} +(1-\alpha)E_{\text{KS}_{X}}^{\text{SR}} + \\
\frac{1}{\varepsilon_{\infty}} E_{_{XX}}^{\text{LR}} + \bigg(1-\frac{1}{\varepsilon_{\infty}} \bigg)E_{\text{KS}_{X}}^{\text{LR}}+ E_{{\text{KS}_{C}}}.
\end{split}
\een
 
For finite systems, the range-separation parameter $\gamma$ was tuned to satisfy the ionization potential (IP) theorem \cite{KSRB12,SEKB10,RBK11,AS14,ZGG19,CMRP18,TLKP20,DFPL10,DFM14,MCRP18,NCFM18}, but a different approach is required for extended systems. This known issue arises due to a delocalization of the KS orbitals, which causes the IP theorem to be trivially satisfied \cite{MCY08,KK15,VESN16,G15}. To address this, the WOT-SRSH functional uses maximally-localized Wannier functions to tune $\gamma$ to satisfy an IP \textit{ansatz} \cite{MW16}. In practice, $\gamma$ is tuned until $\Delta I^{\gamma}=0$ \cite{WOHF21}, where

\ben
\Delta I^{\gamma}=E^{\gamma}_{\text{constr}}[\phi](N-1)-E^{\gamma}(N) + \langle \phi | \hat{H}^{\gamma}_{\text{SRSH}} | \phi \rangle.
\label{eq1}
\een
In Eq.~(\ref{eq1}), the total energy of the system with $N$ electrons, $E^{\gamma}(N)$, is subtracted from the total energy of the system with one electron removed from the Wannier function, $\phi$, $E^{\gamma}_{\text{constr}}[\phi](N-1)$, using a Makov-Payne image charge correction \cite{MP95,WOHF21}. $\langle \phi | \hat{H}^{\gamma}_{\text{SRSH}} | \phi \rangle $
is the expectation value of the energy of a Wannier function, $\phi$, with respect to the SRSH generalized KS Hamiltonian, $\hat{H}^{\gamma}_{\text{SRSH}}$. The crucial aspect here lies in the removal of an electron from the state corresponding to the localized Wannier function, $\phi$, instead of from the delocalized orbital with the highest energy. This removal is enforced by minimizing the total energy of the $N-1$ electron with an added constraint imposed by the equation,
\ben
(\hat{H}_{SRSH} + \lambda |\phi \rangle \langle \phi|) |\psi_{i} \rangle =\epsilon_{i}|\psi_{i}\rangle.
\label{eq.lambda}
\een
Here $\lambda$ acts as a Lagrange multiplier, which at the limit of large $\lambda$ enforces the electron to be fully removed from $\phi$. Eq. (\ref{eq.lambda}) is solved self-consistently for the ($N-1$) electron system, with $\{ \psi_{i } \}$, $\{ \epsilon_{i} \}$ defined as the set of eigenfunctions and eigenvalues corresponding to it. 

To summarize, the WOT-SRSH scheme has four steps. The first step is the calculation of the ion-clamped orientationally-averaged dielectric constant, $\varepsilon_{\infty}$. Second, using a supercell, we obtain a maximally localized Wannier function, $\phi$, with the highest energy (\textit{i.e.} the expectation value of the generalized KS Hamiltonian, with respect to the Wannier function). Third, $\gamma$ is tuned, given a choice of $\alpha$, in the same supercell until $|\Delta I^{\gamma}|<0.02$ eV \cite{WOHF21}. Fourth and finally, we perform SRSH calculations with the tuned $\alpha$ and $\gamma$ values in a primitive cell, to obtain the band gap or any property of interest. Following previous work \cite{WOHF21,OWGC22} we tune in a 2x2x2 supercell using $\Gamma$-point only sampling of the Brillouin zone, and neglecting spin-orbit interactions. Spin-orbit interactions are included in the final calculation (of step 4) and are required to obtain accurate band gaps and band structures for this set of halide double perovskites \cite{FHHA16}. Below, we elaborate and discuss each step in detail. 

\begin{center}
\begin{table}
\begin{tabular}{cccc} \hline \hline
    $F\bar{m}3m$   & $\gamma$ (Bohr\textsuperscript{$-1$}) & $\varepsilon_{\infty}$ & a\textsubscript{lat} (\AA) \\ \hline
    Cs\textsubscript{2}Ag{Bi}Br\textsubscript{6}   & 0.14 & 5.92\cite{BFLB21} & 5.63\cite{MBWW16}\\  
    Cs\textsubscript{2}Ag{Bi}Cl\textsubscript{6} &  0.15 & 4.68\cite{BFLB21} & 5.39\cite{MBWW16}\\ 
    Cs\textsubscript{2}AgSbBr\textsubscript{6}  &  0.14 & 5.96\cite{BFLB21} & 5.58\cite{WDSH19} \\   
    Cs\textsubscript{2}AgSbCl\textsubscript{6} &  0.15 & 4.77\cite{BFLB21} & 5.33\cite{TPCM17} \\ 
    Cs\textsubscript{2}AgTlBr\textsubscript{6} &  0.165 & 3.81 & 5.54\cite{SLSB18} \\ \hline \hline
\end{tabular}
\caption{\label{Table 1} Tuned $\gamma$ values (for $\alpha$ = 0.4), the dielectric constants, $\varepsilon_{\infty}$, and experimental lattice constants, a\textsubscript{$lat$} , used in the calculations.}
\end{table}
\end{center}

\sec{Results}
The tuned $\gamma$ parameters for a fixed choice of $\alpha = 0.4$, as well as the computed orientally averaged dielectric constants $\varepsilon_{\infty} $, obtained for different cubic halide double perovskites are given in Table \ref{Table 1}. Experimental lattice constants at room temperature are used for all calculations\cite{MBWW16,WDSH19,TPCM17}. While Cs$_2$AgBiBr$_6$ transforms into a tetragonal phase at low temperatures exhibiting octahedral tilting and small deviations from the cubic phase \cite{WBSL2021}  we are unaware of low temperature data for the other compounds studied here. The dielectric constants needed for step 1 for the Bi and Sb compounds are taken from Ref.~\cite{BFLB21} where they were obtained from the inverse of the head of dielectric function computed within the random phase approximation using the \texttt{BerkeleyGW} \cite{HL85,HL86,DSSJ12} software package. We use the dielectric constants for the materials of Ref.~\cite{BFLB21} to facilitate comparison with their reported values. The dielectric constant for Cs\textsubscript{2}AgTlBr\textsubscript{6} is computed here in the same manner. Both the $\varepsilon_{\infty}$ and the image charge correction are kept fixed throughout the calculations for consistency, in line with Ref. \cite{WOHF21}. Additional details on all calculations, including a discussion of choice of pseudopotentials, can be found in the Supplementary Information \cite{supplementaryinformation}. 

Following Ref.~\cite{WOHF21}, we select the maximally-localized Wannier function that has the highest expectation energy. As a representative example, in Fig.~\ref{fig wann} we show the isosurface of this Wannier function for Cs$_2$AgSbBr$_6$. We find our calculated Wannier functions to be in line with previously reported orbital contributions to the valence band maximum \cite{BFLB21}. In particular, Fig.~\ref{fig wann} shows a localized \textit{s}-orbital on the Sb atom, with surrounding \textit{p}-orbitals localized on the Cl atom. We note that using a Wannier function with lower energy shifts the gap by $< 0.2$ eV. However, this is not physically motivated, particularly if the Wannier function does not resemble the known orbital contributions of the valence band maximum. 

The tuning curves, namely $\Delta I$ as a function of $\gamma$, are given in Fig. \ref{fig tuning}. 
Fig.~\ref{fig tuning}a shows the tuning curves for all the materials investigated, for $\alpha = 0.4$. The curves are smooth and monotonic, as previously reported for other materials \cite{WOHF21}. The tuned $\gamma$ parameters (for $\alpha$ = 0.4) are found when these curves cross zero, and thus satisfy Eq.~(\ref{eq1}). We emphasize some aspects of the tuned parameters. First, all tuned $\gamma$ values are within $0.02$ Bohr$^{-1}$, consistent with the fact that all compounds have the same $F\bar{m}3m$ cubic phase and exhibit similar chemistry. Second, a larger amount of SR exact exchange is needed compared to other common global hybrids in order to tune the range separation parameter $\gamma$, \textit{i.e.}, no tuned ($\alpha, \gamma$) pairs could be found for $\alpha = 0.25$. This trend has previously been noted in other functionals of different construction (using a distinctly different rationale) \cite{SBC03,GSS03,LFHL00,VJV10}, but also with the WOT-SRSH for lead halide perovskites \cite{OWGC22}. As defined in the generalized KS scheme of RSH functionals, in principle any choice of $\alpha$ maps the exact density equally well \cite{GNGK20}, and thus the choice of $\alpha$ can vary from the traditional value of $0.25$. Further, as previously noted \cite{WOHF21}, when $1/\varepsilon_{\infty}$ $\sim 0.25$, $\Delta I$ becomes independent of $\gamma$, and thus a value different than $0.25$ must be used. 

As shown in Ref.~\cite{WOHF21}, there are many ($\alpha$, $\gamma$) pairs that can satisfy Eq. (\ref{eq1}), such that $|\Delta I^{\gamma} |< 0.02$ eV. In other words, ($\alpha$, $\gamma$) pairs are not unique. However if different, \textit{tuned} ($\alpha$, $\gamma$) pairs are used, the gap (both direct and indirect) varies by less than 0.2 eV, which agrees with Ref.~\cite{WOHF21}. This is shown in Fig.~\ref{fig tuning}b, c for Cs\textsubscript{2}Ag{Bi}Br\textsubscript{6} and  Cs\textsubscript{2}Ag{Bi}Cl\textsubscript{6}, and also demonstrates the stability of the WOT-SRSH functional. 

\begin{center}
\begin{table}
\begin{tabular}{ccccccc}
\hline \hline
{indirect} &
  {PBE} &
  {PBE0} &
  {HSE} &
   {\begin{tabular}[c]{@{}c@{}}$GW$@\\ LDA\end{tabular}\cite{BFLB21}}&
  {\begin{tabular}[c]{@{}c@{}}WOT\\-SRSH\end{tabular}} &
  \begin{tabular}[c]{@{}c@{}}$GW$@WOT\\-SRSH\end{tabular}    \\ \hline 
Cs\textsubscript{2}Ag{Bi}Br\textsubscript{6} & 1.05     & 2.45     & 1.82   & 1.66 & 2.66     & 2.67      \\
Cs\textsubscript{2}AgSbBr\textsubscript{6}   & 0.76     & 2.02     & 1.41   & 1.40 & 2.15     & 2.28       \\
Cs\textsubscript{2}Ag{Bi}Cl\textsubscript{6} & 1.50     &   3.13   & 2.92   & 2.58 & 3.52     & 3.79       \\
Cs\textsubscript{2}AgSbCl\textsubscript{6}   & 1.26     & 2.75     & 2.10   & 2.26 & 3.05     & 3.45        \\
Cs\textsubscript{2}AgTlBr\textsubscript{6}   & --  & --   & -- & -- & -- & -- \\ \hline \hline
\end{tabular}
\caption{\label{Table 2} Indirect (fundamental) gaps of the halide double perovskites, calculated using PBE, PBE0, HSE, $G_0W_0 @$LDA \cite{BFLB21}, WOT-SRSH and $G_{0}W_{0} @$ WOT-SRSH, including spin - orbit coupling effects. All quantities reported in eV.}
\end{table}
\end{center}

\begin{center}
\begin{table}
\begin{tabular}{ccccccc}
\hline \hline
{direct} &
  {PBE} &{PBE0} & {HSE} &  {\begin{tabular}[c]{@{}c@{}}$GW$@\\ LDA\end{tabular}\cite{BFLB21}} &
  {\begin{tabular}[c]{@{}c@{}}WOT\\-SRSH\end{tabular}} &
  \begin{tabular}[c]{@{}c@{}}$GW$@WOT\\-SRSH\end{tabular}   \\ \hline 
Cs\textsubscript{2}Ag{Bi}Br\textsubscript{6}    & 1.76   & 3.13   & 2.48   & 2.41   & 3.31 & 3.38  \\
Cs\textsubscript{2}AgSbBr\textsubscript{6}     & 1.93   & 3.22   & 2.65    & 2.73   & 3.32 & 3.69  \\
Cs\textsubscript{2}Ag{Bi}Cl\textsubscript{6}  & 2.05   & 3.64   & 2.79    & 2.98   & 4.03 & 4.27  \\
Cs\textsubscript{2}AgSbCl\textsubscript{6}   & 2.37   & 3.86    & 3.18    & 3.43   & 4.10 & 4.56   \\
Cs\textsubscript{2}AgTlBr\textsubscript{6}    & 0.00    & 0.92   & 0.17    & --  & 1.36 & 1.11    \\ \hline \hline
\end{tabular}
\caption{\label{Table 3} Direct gaps of the halide double perovskites, calculated using PBE, PBE0, HSE, $G_0W_0 @$LDA \cite{BFLB21}, WOT-SRSH and $G_{0}W_{0} @$ WOT-SRSH, including spin - orbit coupling effects. All quantities reported in eV.}
\end{table}
\end{center}
Tables \ref{Table 2}, and \ref{Table 3} summarize the indirect and direct gaps computed using WOT-SRSH and $G_0W_0$@WOT-SRSH, and include for comparison results computed with PBE \cite{PBE96}, PBE0 \cite{PBE0} and HSE \cite{HSE06} functionals, as well as $G_0W_0$@LDA \cite{BFLB21}. With the exception of Cs\textsubscript{2}AgTlBr\textsubscript{6}, which has a direct fundamental gap at $\Gamma$, all materials investigated have an indirect fundamental gap with the valence band maximum at the high symmetry point $X$ in the Brillouin zone, and the conduction band minimum at $L$ . As reported previously, the PBE gaps are smaller than those predicted by other methods by at least $\sim 1$ eV; also consistent with prior calculations, PBE erroneuously predicts the Tl compound to be metallic. While direct and indirect band gaps computed with PBE0, HSE, and WOT-SRSH exhibit similar trends for the five halide double perovskites studied here, the different functionals predict gaps that are significantly different quantitatively. Notably, while the HSE gaps are all larger than those computed with PBE -- by close to a factor of two for Cs$_2$AgSbBr$_6$ and Cs$_2$AgBiCl$_6$ -- the PBE0 gaps can be up to 50\% larger than HSE. The significant quantitative differences with WOT-SRSH found here for different  hybrid functionals (with the exception of PBE0), and the sensitivity to starting point for $G_0W_0$ have been discussed previously, where it was noted that a ``one-size-fits-all'' functional for predicting band gaps is lacking for double and single halide perovskites \cite{LRN19}. Moreover, the WOT-SRSH gaps are computed to be somewhat larger than the PBE0 gaps. Although they are in general within 0.2 eV of PBE0 in most cases, the exceptions are Cs$_2$AgBiCl$_6$ and Cs$_2$AgTlBr$_6$, where the WOT-SRSH gaps are noticeably larger compared to PBE0 -- a clear contrast between the two functionals. While $G_0W_0$ corrections to the LDA gaps lead to values aligned with HSE, they are significantly smaller than $G_0W_0$ corrections to WOT-SRSH. This is possibly due to LDA severely underestimating the gaps for these systems, and thus affecting the $G_0W_0$ results, when used as a starting point. We further note that the $G_0W_0$ values obtained with WOT-SRSH are much larger when compared to $G_0W_0$@LDA, and more consistent with the uncorrected WOT-SRSH gaps. This is in agreement with prior work finding small $GW$ corrections to WOT-SRSH gaps of semiconductors and insulators \cite{GHFS22}. Consistent with prior reports \cite{BFLB21,LASLK23}, we find the Cl compounds to have a larger gap than Br compounds (for a fixed Bi or Sb), associated with the fact that the Cl compounds have smaller volumes and therefore greater hybridization. Likewise we find the Bi compounds have smaller gaps than those with Sb likely due to Bi 6s and 6p contributions to the band edges being more delocalized than their Sb 5s and 5p counterparts.

Despite experimental interest in the halide double perovskites explored here, we are not aware of direct measurements of the fundamental gaps of these compounds, rendering a direct comparison to experiment challenging. Reports of measured band gaps in the literature are often derived from optical data, usually taken at room temperature, relevant to device operating conditions. Optical gaps are by definition smaller than fundamental gaps, with the difference being the exciton binding energy. While previous calculations of these materials (excluding Cs\textsubscript{2}AgTlBr\textsubscript{6}) report resonant excitons corresponding to direct optical transitions at $X$ and $L$  (with relatively large exciton binding energies of $\sim 0.2- 0.4$ eV \cite{BFLB21}), prior band gaps extracted from measured optical absorption spectra were obtained from Tauc fits of spectra that do not account for electron-hole interactions (likely an acceptable approximation for these compounds \cite{SLSB18}). In particular, all indirect gaps reported from experiments were extrapolated from Tauc plots \cite{WBSL2021,WDSH19,TPCM17}. For the reported experimental gap \cite{SLSB18,DSWF20} of Cs\textsubscript{2}AgTlBr\textsubscript{6}, a Kubelka-Munk transformation was used to determine the gap, which also neglects electron-hole interactions. Further, distinct experimental reports for the indirect gap of Cs\textsubscript{2}Ag{Bi}Br\textsubscript{6} differ by $0.3$ eV \cite{VFHS16,FHHA16,SHLK16}. 

In what follows, we compare our WOT-SRSH calculations with experimental optical spectra for Cs\textsubscript{2}Ag{Bi}Br\textsubscript{6}, which accounts for excitonic effects \cite{WBSL2021} via an Elliot fit that captures electron-hole interactions assuming hydrogenic Wannier-Mott excitons \cite{E57,W37}. Comparing the reported $\sim 3$ eV direct gap of Cs\textsubscript{2}Ag{Bi}Br\textsubscript{6} extracted from this experiment, we find a value within $\sim 0.3$ eV of our WOT-SRSH direct gap. While this difference is larger than those previously found for simpler systems \cite{WOHF21, GHFS22}, it lends strong support to the claim of quantitative improvement of non-empirical WOT-SRSH and $G_0W_0$@WOT-SRSH gap values, which are systematically larger than those we obtain with the hybrid functionals PBE0 and HSE. We also observe that $G_0W_0$ calculations using the WOT-SRSH eigensystem yield a minimal correction of 0.07 eV for the direct gap and 0.01 eV for the indirect gap. Such a small change is in line with previous work \cite{GHFS22} and also suggests that the WOT-SRSH functional is accurate and predictive for these materials. From our WOT-SRSH calculations and single-shot $G_0W_0$ calculations on top of WOT-SRSH, we can conclude that nonempirical WOT-SRSH functionals are more predictive than other functionals to date, providing values and trends within the expected range for the halide double perovskites and exhibiting similar performance as for other systems \cite{OWGC22}. 

\begin{figure}
    \centering
    \subfigure(a){\includegraphics[width=0.9\linewidth]{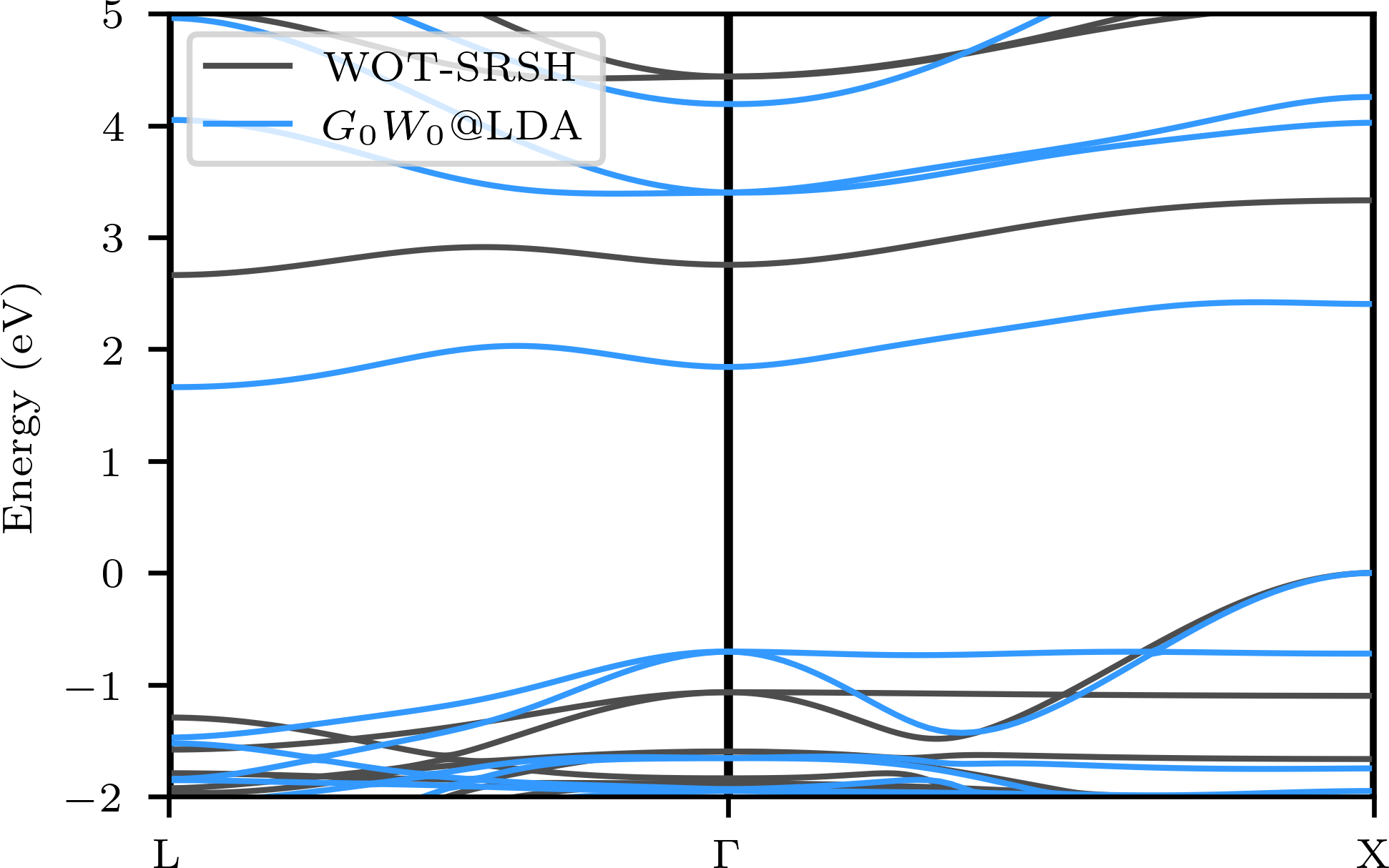}} 
    \subfigure(b){\includegraphics[width=0.9\linewidth]{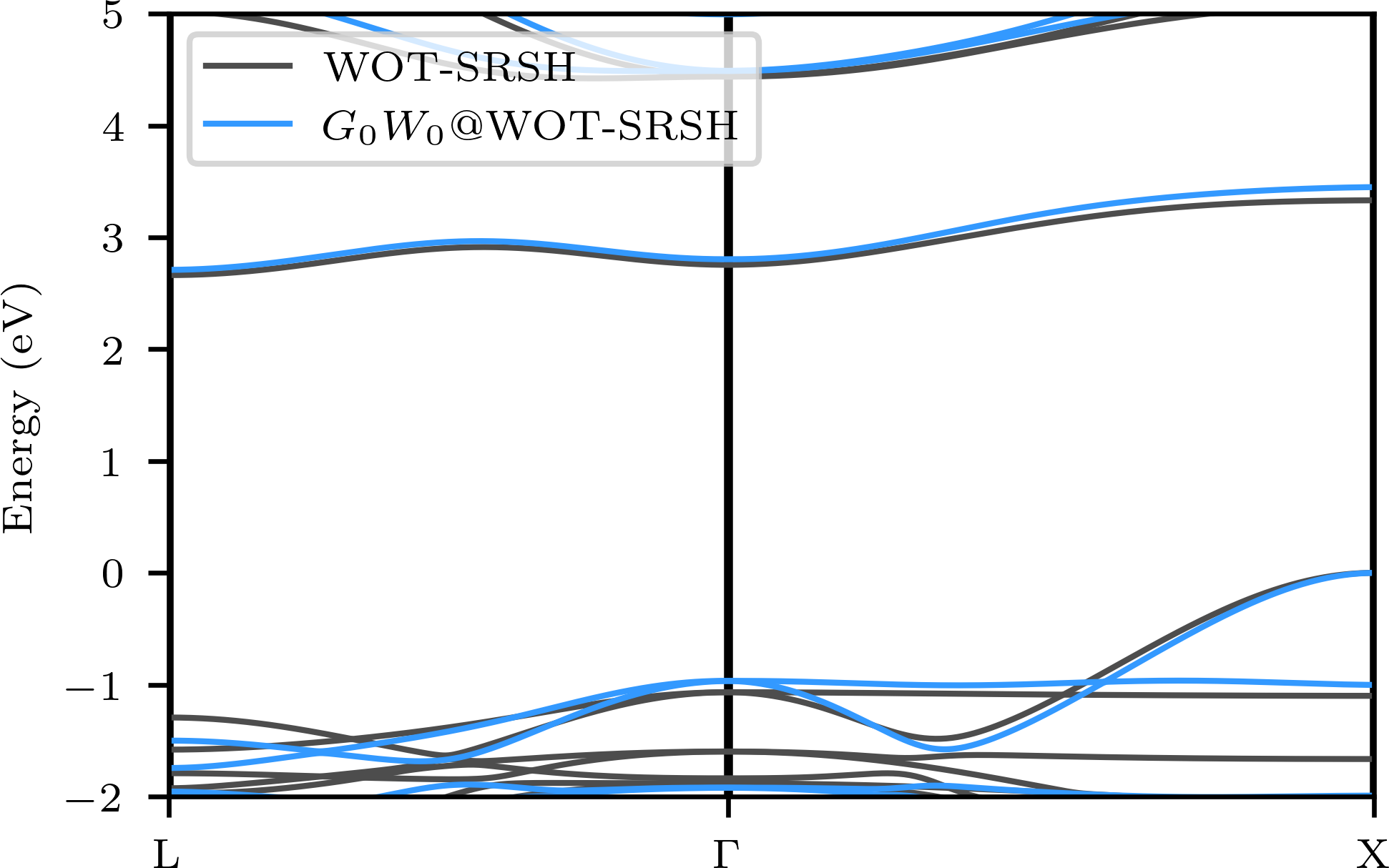}}
    \caption{Band structures of Cs\textsubscript{2}Ag{Bi}Br\textsubscript{6}, comparing (a) WOT-SRSH (black) with previous $G_{0}W_{0}$@LDA (blue) calculations \cite{BFLB21}, and (b) WOT-SRSH (black) with  $G_{0}W_{0}$@WOT-SRSH (blue).}
    \label{fig 5}
\end{figure}

We can also compare the previously-reported $G_0W_0$@LDA Cs\textsubscript{2}Ag{Bi}Cl\textsubscript{6} band structure \cite{BFLB21} with that computed here with the WOT-SRSH functional, as shown in Fig. \ref{fig 5}. As shown in Fig.~\ref{fig 5}a, the Wannier-interpolated WOT-SRSH band structure agrees qualitatively (\textit{i.e.}, band dispersion) with the previous $G_0W_0$@LDA result, with the main difference being that the magnitude of the WOT-SRSH gap is larger. Likewise, Fig. \ref{fig 5}b compares the WOT-SRSH band structure along with $G_0W_0$@WOT-SRSH, which again depicts agreement in band shape, but also in the magnitude of the gap itself. This agrees with the results of Tables ~\ref{Table 1} and \ref{Table 2}, and the trends with methods are also consistent with reported band structures of the simpler Pb-based halide perovskites \cite{OWGC22}.

\begin{figure}[!h]
\includegraphics[width=0.95\linewidth]{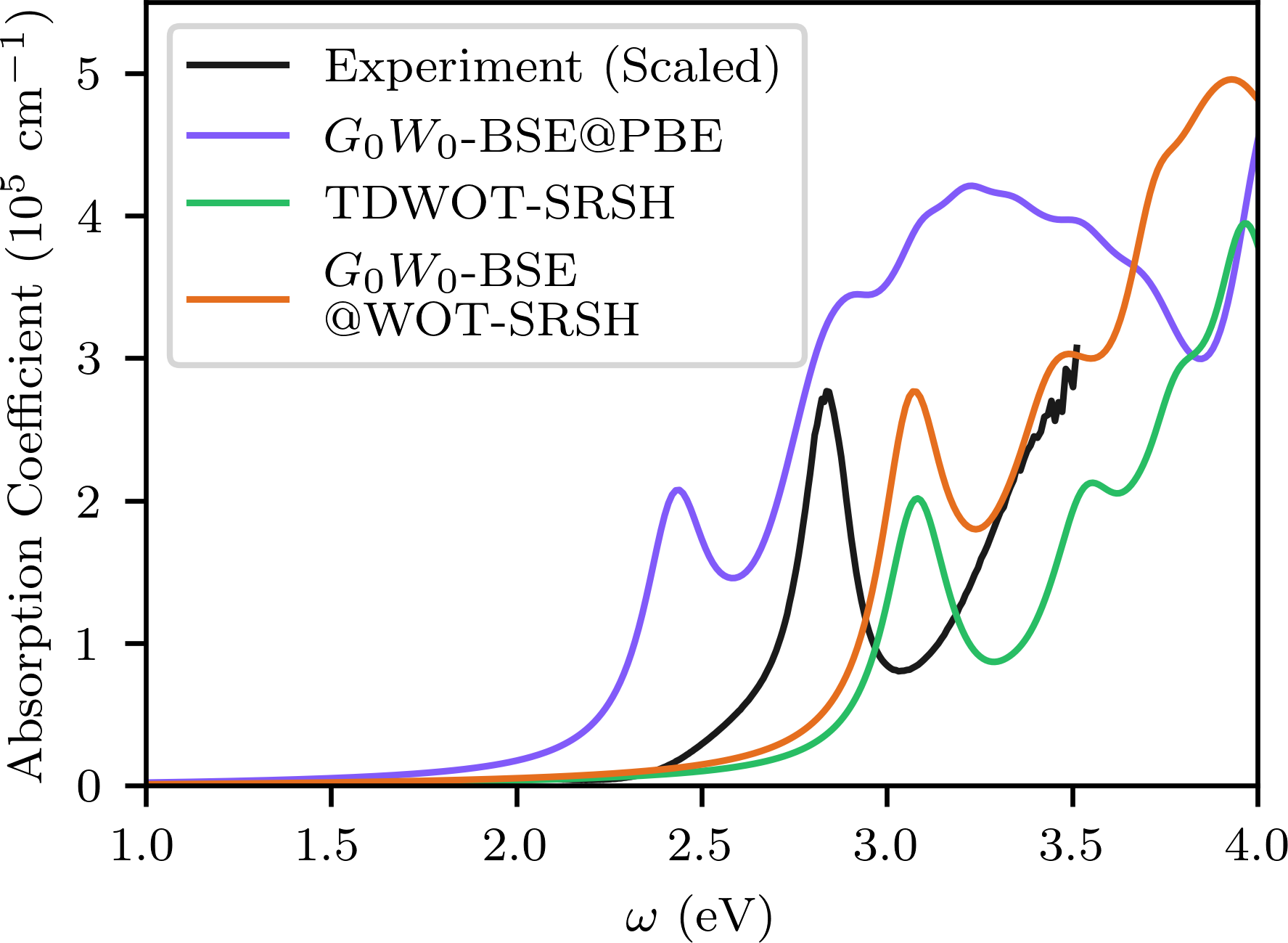}
\caption{Optical absorption coefficient $\nicefrac{\omega \varepsilon_2}{c n}$ for Cs\textsubscript{2}Ag{Bi}Br\textsubscript{6}, where $\varepsilon_2$ is the imaginary part of the dielectric function, $c$ is the speed of light, and $n$ is real part of the index of refraction. The $G_0W_0$-BSE@PBE denoted in purple, experiment data from Ref.~\cite{LMBW20} in black, TDWOT-SRSH in green, and $G_0W_0$-BSE@WOT-SRSH in orange.   Original experiment from \cite{LMBW20} reports absorbance (unitless) as a function of wave length. To compare with our calculation of the absorption coefficient (in units of $cm^{-1}$) as a function of photon energy, we have transformed the energy axis in eV and have also rescaled the experimental data such that the height of the excitonic peak is the same as from $G_0W_0$-BSE@WOT-SRSH.}
\label{fig abs}
\end{figure}

Finally, in Fig. \ref{fig abs} we plot the linear optical absorption spectra of Cs\textsubscript{2}Ag{Bi}Br\textsubscript{6} using time-dependent WOT-SRSH (denoted as TDWOT-SRSH), $G_0W_0$-BSE@PBE, and $G_0W_0$-BSE@WOT-SRSH, comparing to re-scaled experimental data taken from Ref.~\cite{LMBW20}. In Ref. \cite{WBSL2021}, excitonic effects were considered in an Elliot fit of the Cs\textsubscript{2}Ag{Bi}Br\textsubscript{6} absorption spectrum, and an exciton binding energy of approximately 200 meV was reported at room temperature. The $G_0W_0$-BSE@LDA spectra underestimates the experimental absorption onset by $\sim 0.4$ eV, as previously noted \cite{BFLB21}, while those obtained using WOT-SRSH either with $G_0W_0$-BSE or TDDFT overestimate it by a smaller amount, $\sim0.2$ eV. The WOT-SRSH overestimate of the onset by $0.2$ eV is a clear improvement over prior calculations in comparison with these room temperature experiments, as our calculations neglect the effects of lattice vibrations on the peak energies and line shapes. As we have discussed in our analysis of other compounds in prior work \cite{WOHF21,OWGC22,OG2023}, such effects will tend to redshift peak energies, and thus we would expect to overestimate peak energies before introducing phonon effects. Moreover, we note there are additional uncertainties associated with Elliot fit, given the low-lying excitons have been reported to be nonhydrogenic \cite{BCFL23}. Finally, both spectra obtained using WOT-SRSH reproduce the line shape associated with the first valley in the experimental spectrum, while the spectra derived from a PBE starting point does not.  

\sec{Conclusions}
We have applied the WOT-SRSH functional to a set of halide double perovskites, a class of materials for which the prediction of the fundamental and direct gaps with any known approach has been a challenge. Previous calculations with both hybrid XC functionals and single shot $G_0W_0$ approaches have fallen short of predicting a consistent set of gaps. Here we find that WOT-SRSH predicts the gaps in a consistent manner when compared to prior $G_0W_0$@LDA calculations. We find our WOT-SRSH calculations  agree best with previous experimental results when accounting for excitonic effects (and modulo the effects of lattice vibrations), and that the improvements to predicted band gaps over other approaches is consistent with previously investigated materials \cite{WOHF21,OWGC22,OG2023.} We also find agreement with previously reported band structures using WOT-SRSH. Finally, for Cs\textsubscript{2}Ag{Bi}Br\textsubscript{6}, we find that WOT-SRSH can improve on predicted optical absorption spectra when used either as a starting point to TDDFT calculations or for $G_0W_0$-BSE.

\sec{Acknowledgments}
The authors acknowledge Giulia Longo (Northumbria U.) and Laura Herz (Oxford U.) for for sharing the data published in Ref.~\cite{LMBW20}. L.K. acknowledges support from the Arieh and Mintzy Katzman Professoral Chair and the Helen and Martin Kimmel Award for Innovative investigation. M.R.F. acknowledges support from the UK Engineering and Physical Sciences Research Council
(EPSRC), Grant No. EP/V010840/1. This work is funded through NSF–Binational
Science Foundation Grant No.~DMR-2015991 and by the Israel Science Foundation. Computational resources were provided by the NSF funded XSEDE (now ACCESS) program, through supercomputer Stampede2 at the
Texas Advanced Computing Center (TACC) through the allocation TG-DMR190070. 

\clearpage
\bibliography{double_perovskites_citations}

\begin{thebibliography}{78}%
\makeatletter
\providecommand \@ifxundefined [1]{%
 \@ifx{#1\undefined}
}%
\providecommand \@ifnum [1]{%
 \ifnum #1\expandafter \@firstoftwo
 \else \expandafter \@secondoftwo
 \fi
}%
\providecommand \@ifx [1]{%
 \ifx #1\expandafter \@firstoftwo
 \else \expandafter \@secondoftwo
 \fi
}%
\providecommand \natexlab [1]{#1}%
\providecommand \enquote  [1]{``#1''}%
\providecommand \bibnamefont  [1]{#1}%
\providecommand \bibfnamefont [1]{#1}%
\providecommand \citenamefont [1]{#1}%
\providecommand \href@noop [0]{\@secondoftwo}%
\providecommand \href [0]{\begingroup \@sanitize@url \@href}%
\providecommand \@href[1]{\@@startlink{#1}\@@href}%
\providecommand \@@href[1]{\endgroup#1\@@endlink}%
\providecommand \@sanitize@url [0]{\catcode `\\12\catcode `\$12\catcode
  `\&12\catcode `\#12\catcode `\^12\catcode `\_12\catcode `\%12\relax}%
\providecommand \@@startlink[1]{}%
\providecommand \@@endlink[0]{}%
\providecommand \url  [0]{\begingroup\@sanitize@url \@url }%
\providecommand \@url [1]{\endgroup\@href {#1}{\urlprefix }}%
\providecommand \urlprefix  [0]{URL }%
\providecommand \Eprint [0]{\href }%
\providecommand \doibase [0]{http://dx.doi.org/}%
\providecommand \selectlanguage [0]{\@gobble}%
\providecommand \bibinfo  [0]{\@secondoftwo}%
\providecommand \bibfield  [0]{\@secondoftwo}%
\providecommand \translation [1]{[#1]}%
\providecommand \BibitemOpen [0]{}%
\providecommand \bibitemStop [0]{}%
\providecommand \bibitemNoStop [0]{.\EOS\space}%
\providecommand \EOS [0]{\spacefactor3000\relax}%
\providecommand \BibitemShut  [1]{\csname bibitem#1\endcsname}%
\let\auto@bib@innerbib\@empty
\bibitem [{\citenamefont {McClure}\ \emph {et~al.}(2016)\citenamefont
  {McClure}, \citenamefont {Ball}, \citenamefont {Windl},\ and\ \citenamefont
  {Woodward}}]{MBWW16}%
  \BibitemOpen
  \bibfield  {author} {\bibinfo {author} {\bibfnamefont {E.~T.}\ \bibnamefont
  {McClure}}, \bibinfo {author} {\bibfnamefont {M.~R.}\ \bibnamefont {Ball}},
  \bibinfo {author} {\bibfnamefont {W.}~\bibnamefont {Windl}}, \ and\ \bibinfo
  {author} {\bibfnamefont {P.~M.}\ \bibnamefont {Woodward}},\ }\href@noop {}
  {\bibfield  {journal} {\bibinfo  {journal} {Chem. Mater.}\ }\textbf {\bibinfo
  {volume} {28}},\ \bibinfo {pages} {1348} (\bibinfo {year}
  {2016})}\BibitemShut {NoStop}%
\bibitem [{\citenamefont {Volonakis}\ \emph {et~al.}(2016)\citenamefont
  {Volonakis}, \citenamefont {Filip}, \citenamefont {Haghighirad},
  \citenamefont {Sakai}, \citenamefont {Wenger}, \citenamefont {Snaith},\ and\
  \citenamefont {Giustino}}]{VFHS16}%
  \BibitemOpen
  \bibfield  {author} {\bibinfo {author} {\bibfnamefont {G.}~\bibnamefont
  {Volonakis}}, \bibinfo {author} {\bibfnamefont {M.~R.}\ \bibnamefont
  {Filip}}, \bibinfo {author} {\bibfnamefont {A.~A.}\ \bibnamefont
  {Haghighirad}}, \bibinfo {author} {\bibfnamefont {N.}~\bibnamefont {Sakai}},
  \bibinfo {author} {\bibfnamefont {B.}~\bibnamefont {Wenger}}, \bibinfo
  {author} {\bibfnamefont {H.~J.}\ \bibnamefont {Snaith}}, \ and\ \bibinfo
  {author} {\bibfnamefont {F.}~\bibnamefont {Giustino}},\ }\href {\doibase
  10.1021/acs.jpclett.6b00376} {\bibfield  {journal} {\bibinfo  {journal} {J.
  Phys. Chem. Lett.}\ }\textbf {\bibinfo {volume} {7}},\ \bibinfo {pages}
  {1254--1259} (\bibinfo {year} {2016})}\BibitemShut {NoStop}%
\bibitem [{\citenamefont {Wei}\ \emph {et~al.}(2016)\citenamefont {Wei},
  \citenamefont {Deng}, \citenamefont {Sun}, \citenamefont {Xie}, \citenamefont
  {Kieslich}, \citenamefont {Evans}, \citenamefont {Carpenter}, \citenamefont
  {Bristowe},\ and\ \citenamefont {Cheetham}}]{WDSX16}%
  \BibitemOpen
  \bibfield  {author} {\bibinfo {author} {\bibfnamefont {F.}~\bibnamefont
  {Wei}}, \bibinfo {author} {\bibfnamefont {Z.}~\bibnamefont {Deng}}, \bibinfo
  {author} {\bibfnamefont {S.}~\bibnamefont {Sun}}, \bibinfo {author}
  {\bibfnamefont {F.}~\bibnamefont {Xie}}, \bibinfo {author} {\bibfnamefont
  {G.}~\bibnamefont {Kieslich}}, \bibinfo {author} {\bibfnamefont {D.~M.}\
  \bibnamefont {Evans}}, \bibinfo {author} {\bibfnamefont {M.~A.}\ \bibnamefont
  {Carpenter}}, \bibinfo {author} {\bibfnamefont {P.~D.}\ \bibnamefont
  {Bristowe}}, \ and\ \bibinfo {author} {\bibfnamefont {A.~K.}\ \bibnamefont
  {Cheetham}},\ }\href {\doibase 10.1039/C6MH00053C} {\bibfield  {journal}
  {\bibinfo  {journal} {Mater. Horiz.}\ }\textbf {\bibinfo {volume} {3}},\
  \bibinfo {pages} {328--332} (\bibinfo {year} {2016})}\BibitemShut {NoStop}%
\bibitem [{\citenamefont {Volonakis}\ \emph {et~al.}(2017)\citenamefont
  {Volonakis}, \citenamefont {Haghighirad}, \citenamefont {Milot},
  \citenamefont {Sio}, \citenamefont {Filip}, \citenamefont {Wenger},
  \citenamefont {Johnston}, \citenamefont {Herz}, \citenamefont {Snaith},\ and\
  \citenamefont {Giustino}}]{VHMS17}%
  \BibitemOpen
  \bibfield  {author} {\bibinfo {author} {\bibfnamefont {G.}~\bibnamefont
  {Volonakis}}, \bibinfo {author} {\bibfnamefont {A.~A.}\ \bibnamefont
  {Haghighirad}}, \bibinfo {author} {\bibfnamefont {R.~L.}\ \bibnamefont
  {Milot}}, \bibinfo {author} {\bibfnamefont {W.~H.}\ \bibnamefont {Sio}},
  \bibinfo {author} {\bibfnamefont {M.~R.}\ \bibnamefont {Filip}}, \bibinfo
  {author} {\bibfnamefont {B.}~\bibnamefont {Wenger}}, \bibinfo {author}
  {\bibfnamefont {M.~B.}\ \bibnamefont {Johnston}}, \bibinfo {author}
  {\bibfnamefont {L.~M.}\ \bibnamefont {Herz}}, \bibinfo {author}
  {\bibfnamefont {H.~J.}\ \bibnamefont {Snaith}}, \ and\ \bibinfo {author}
  {\bibfnamefont {F.}~\bibnamefont {Giustino}},\ }\href {\doibase
  10.1021/acs.jpclett.6b02682} {\bibfield  {journal} {\bibinfo  {journal} {J.
  Phys. Chem. Lett.}\ }\textbf {\bibinfo {volume} {8}},\ \bibinfo {pages}
  {772--778} (\bibinfo {year} {2017})}\BibitemShut {NoStop}%
\bibitem [{\citenamefont {Slavney}\ \emph {et~al.}(2018)\citenamefont
  {Slavney}, \citenamefont {Leppert}, \citenamefont {SaldivarValdes},
  \citenamefont {Bartesaghi}, \citenamefont {Savenije}, \citenamefont
  {Neaton},\ and\ \citenamefont {Karunadasa}}]{SLSB18}%
  \BibitemOpen
  \bibfield  {author} {\bibinfo {author} {\bibfnamefont {A.~H.}\ \bibnamefont
  {Slavney}}, \bibinfo {author} {\bibfnamefont {L.}~\bibnamefont {Leppert}},
  \bibinfo {author} {\bibfnamefont {A.}~\bibnamefont {SaldivarValdes}},
  \bibinfo {author} {\bibfnamefont {D.}~\bibnamefont {Bartesaghi}}, \bibinfo
  {author} {\bibfnamefont {T.~J.}\ \bibnamefont {Savenije}}, \bibinfo {author}
  {\bibfnamefont {J.~B.}\ \bibnamefont {Neaton}}, \ and\ \bibinfo {author}
  {\bibfnamefont {H.~I.}\ \bibnamefont {Karunadasa}},\ }\href {\doibase
  https://doi.org/10.1002/anie.201807421} {\bibfield  {journal} {\bibinfo
  {journal} {Angew. Chem. Int. Ed.}\ }\textbf {\bibinfo {volume} {57}},\
  \bibinfo {pages} {12765--12770} (\bibinfo {year} {2018})}\BibitemShut
  {NoStop}%
\bibitem [{\citenamefont {Faber}\ \emph {et~al.}(2016)\citenamefont {Faber},
  \citenamefont {Lindmaa}, \citenamefont {von Lilienfeld},\ and\ \citenamefont
  {Armiento}}]{FLVA16}%
  \BibitemOpen
  \bibfield  {author} {\bibinfo {author} {\bibfnamefont {F.~A.}\ \bibnamefont
  {Faber}}, \bibinfo {author} {\bibfnamefont {A.}~\bibnamefont {Lindmaa}},
  \bibinfo {author} {\bibfnamefont {O.~A.}\ \bibnamefont {von Lilienfeld}}, \
  and\ \bibinfo {author} {\bibfnamefont {R.}~\bibnamefont {Armiento}},\ }\href
  {\doibase 10.1103/PhysRevLett.117.135502} {\bibfield  {journal} {\bibinfo
  {journal} {Phys. Rev. Lett.}\ }\textbf {\bibinfo {volume} {117}},\ \bibinfo
  {pages} {135502} (\bibinfo {year} {2016})}\BibitemShut {NoStop}%
\bibitem [{\citenamefont {Filip}\ and\ \citenamefont {Giustino}(2018)}]{FG18}%
  \BibitemOpen
  \bibfield  {author} {\bibinfo {author} {\bibfnamefont {M.~R.}\ \bibnamefont
  {Filip}}\ and\ \bibinfo {author} {\bibfnamefont {F.}~\bibnamefont
  {Giustino}},\ }\href {\doibase 10.1073/pnas.1719179115} {\bibfield  {journal}
  {\bibinfo  {journal} {PNAS}\ }\textbf {\bibinfo {volume} {115}},\ \bibinfo
  {pages} {5397--5402} (\bibinfo {year} {2018})}\BibitemShut {NoStop}%
\bibitem [{\citenamefont {Slavney}\ \emph
  {et~al.}(2016{\natexlab{a}})\citenamefont {Slavney}, \citenamefont {Hu},
  \citenamefont {Lindenberg},\ and\ \citenamefont {Karunadasa}}]{SHLK16}%
  \BibitemOpen
  \bibfield  {author} {\bibinfo {author} {\bibfnamefont {A.~H.}\ \bibnamefont
  {Slavney}}, \bibinfo {author} {\bibfnamefont {T.}~\bibnamefont {Hu}},
  \bibinfo {author} {\bibfnamefont {A.~M.}\ \bibnamefont {Lindenberg}}, \ and\
  \bibinfo {author} {\bibfnamefont {H.~I.}\ \bibnamefont {Karunadasa}},\ }\href
  {\doibase 10.1021/jacs.5b13294} {\bibfield  {journal} {\bibinfo  {journal}
  {J. Am. Chem. Soc.}\ }\textbf {\bibinfo {volume} {138}},\ \bibinfo {pages}
  {2138--2141} (\bibinfo {year} {2016}{\natexlab{a}})}\BibitemShut {NoStop}%
\bibitem [{\citenamefont {Deng}\ \emph {et~al.}(2016)\citenamefont {Deng},
  \citenamefont {Wei}, \citenamefont {Sun}, \citenamefont {Kieslich},
  \citenamefont {Cheetham},\ and\ \citenamefont {Bristowe}}]{DWSK16}%
  \BibitemOpen
  \bibfield  {author} {\bibinfo {author} {\bibfnamefont {Z.}~\bibnamefont
  {Deng}}, \bibinfo {author} {\bibfnamefont {F.}~\bibnamefont {Wei}}, \bibinfo
  {author} {\bibfnamefont {S.}~\bibnamefont {Sun}}, \bibinfo {author}
  {\bibfnamefont {G.}~\bibnamefont {Kieslich}}, \bibinfo {author}
  {\bibfnamefont {A.~K.}\ \bibnamefont {Cheetham}}, \ and\ \bibinfo {author}
  {\bibfnamefont {P.~D.}\ \bibnamefont {Bristowe}},\ }\href
  {http://dx.doi.org/10.1039/C6TA05817E} {\bibfield  {journal} {\bibinfo
  {journal} {J. Mater. Chem. A}\ }\textbf {\bibinfo {volume} {4}},\ \bibinfo
  {pages} {12025--12029} (\bibinfo {year} {2016})}\BibitemShut {NoStop}%
\bibitem [{\citenamefont {Debbichi}\ \emph {et~al.}(2018)\citenamefont
  {Debbichi}, \citenamefont {Lee}, \citenamefont {Cho}, \citenamefont {Rappe},
  \citenamefont {Hong}, \citenamefont {Jang},\ and\ \citenamefont
  {Kim}}]{DLCR18}%
  \BibitemOpen
  \bibfield  {author} {\bibinfo {author} {\bibfnamefont {L.}~\bibnamefont
  {Debbichi}}, \bibinfo {author} {\bibfnamefont {S.}~\bibnamefont {Lee}},
  \bibinfo {author} {\bibfnamefont {H.}~\bibnamefont {Cho}}, \bibinfo {author}
  {\bibfnamefont {A.~M.}\ \bibnamefont {Rappe}}, \bibinfo {author}
  {\bibfnamefont {K.-H.}\ \bibnamefont {Hong}}, \bibinfo {author}
  {\bibfnamefont {M.~S.}\ \bibnamefont {Jang}}, \ and\ \bibinfo {author}
  {\bibfnamefont {H.}~\bibnamefont {Kim}},\ }\href {\doibase
  https://doi.org/10.1002/adma.201707001} {\ \textbf {\bibinfo {volume} {30}},\
  \bibinfo {pages} {1707001} (\bibinfo {year} {2018})}\BibitemShut {NoStop}%
\bibitem [{\citenamefont {Yang}\ \emph {et~al.}(2018)\citenamefont {Yang},
  \citenamefont {Chen}, \citenamefont {Yang}, \citenamefont {Hong},
  \citenamefont {Sun}, \citenamefont {Han}, \citenamefont {Pullerits},
  \citenamefont {Deng},\ and\ \citenamefont {Han}}]{YCYH18}%
  \BibitemOpen
  \bibfield  {author} {\bibinfo {author} {\bibfnamefont {B.}~\bibnamefont
  {Yang}}, \bibinfo {author} {\bibfnamefont {J.}~\bibnamefont {Chen}}, \bibinfo
  {author} {\bibfnamefont {S.}~\bibnamefont {Yang}}, \bibinfo {author}
  {\bibfnamefont {F.}~\bibnamefont {Hong}}, \bibinfo {author} {\bibfnamefont
  {L.}~\bibnamefont {Sun}}, \bibinfo {author} {\bibfnamefont {P.}~\bibnamefont
  {Han}}, \bibinfo {author} {\bibfnamefont {P.~T.}\ \bibnamefont {Pullerits}},
  \bibinfo {author} {\bibfnamefont {P.~W.}\ \bibnamefont {Deng}}, \ and\
  \bibinfo {author} {\bibfnamefont {P.~K.}\ \bibnamefont {Han}},\ }\href
  {https://onlinelibrary.wiley.com/doi/abs/10.1002/anie.201800660} {\bibfield
  {journal} {\bibinfo  {journal} {Angew. Chem., Int. Ed.}\ }\textbf {\bibinfo
  {volume} {57}},\ \bibinfo {pages} {5359} (\bibinfo {year}
  {2018})}\BibitemShut {NoStop}%
\bibitem [{\citenamefont {Biega}\ \emph {et~al.}(2021)\citenamefont {Biega},
  \citenamefont {Filip}, \citenamefont {Leppert},\ and\ \citenamefont
  {Neaton}}]{BFLB21}%
  \BibitemOpen
  \bibfield  {author} {\bibinfo {author} {\bibfnamefont {R.}~\bibnamefont
  {Biega}}, \bibinfo {author} {\bibfnamefont {M.~R.}\ \bibnamefont {Filip}},
  \bibinfo {author} {\bibfnamefont {L.}~\bibnamefont {Leppert}}, \ and\
  \bibinfo {author} {\bibfnamefont {J.~B.}\ \bibnamefont {Neaton}},\ }\href
  {\doibase 10.1021/acs.jpclett.0c03579} {\bibfield  {journal} {\bibinfo
  {journal} {J. Phys. Chem. Lett.}\ }\textbf {\bibinfo {volume} {12}},\
  \bibinfo {pages} {2057--2063} (\bibinfo {year} {2021})}\BibitemShut {NoStop}%
\bibitem [{\citenamefont {Hohenberg}\ and\ \citenamefont {Kohn}(1964)}]{HK64}%
  \BibitemOpen
  \bibfield  {author} {\bibinfo {author} {\bibfnamefont {P.}~\bibnamefont
  {Hohenberg}}\ and\ \bibinfo {author} {\bibfnamefont {W.}~\bibnamefont
  {Kohn}},\ }\href {\doibase 10.1103/PhysRev.136.B864} {\bibfield  {journal}
  {\bibinfo  {journal} {Phys. Rev.}\ }\textbf {\bibinfo {volume} {136}},\
  \bibinfo {pages} {B864--B871} (\bibinfo {year} {1964})}\BibitemShut {NoStop}%
\bibitem [{\citenamefont {Kohn}\ and\ \citenamefont {Sham}(1965)}]{KS65}%
  \BibitemOpen
  \bibfield  {author} {\bibinfo {author} {\bibfnamefont {W.}~\bibnamefont
  {Kohn}}\ and\ \bibinfo {author} {\bibfnamefont {L.~J.}\ \bibnamefont
  {Sham}},\ }\href {\doibase 10.1103/PhysRev.140.A1133} {\bibfield  {journal}
  {\bibinfo  {journal} {Phys. Rev.}\ }\textbf {\bibinfo {volume} {140}},\
  \bibinfo {pages} {A1133--A1138} (\bibinfo {year} {1965})}\BibitemShut
  {NoStop}%
\bibitem [{\citenamefont {Rohlfing}\ and\ \citenamefont {Louie}(2000)}]{RL00}%
  \BibitemOpen
  \bibfield  {author} {\bibinfo {author} {\bibfnamefont {M.}~\bibnamefont
  {Rohlfing}}\ and\ \bibinfo {author} {\bibfnamefont {S.~G.}\ \bibnamefont
  {Louie}},\ }\href {\doibase 10.1103/PhysRevB.62.4927} {\bibfield  {journal}
  {\bibinfo  {journal} {Phys. Rev. B}\ }\textbf {\bibinfo {volume} {62}},\
  \bibinfo {pages} {4927--4944} (\bibinfo {year} {2000})}\BibitemShut {NoStop}%
\bibitem [{\citenamefont {Albrecht}\ \emph {et~al.}(1998)\citenamefont
  {Albrecht}, \citenamefont {Reining}, \citenamefont {Del~Sole},\ and\
  \citenamefont {Onida}}]{ARDO98}%
  \BibitemOpen
  \bibfield  {author} {\bibinfo {author} {\bibfnamefont {S.}~\bibnamefont
  {Albrecht}}, \bibinfo {author} {\bibfnamefont {L.}~\bibnamefont {Reining}},
  \bibinfo {author} {\bibfnamefont {R.}~\bibnamefont {Del~Sole}}, \ and\
  \bibinfo {author} {\bibfnamefont {G.}~\bibnamefont {Onida}},\ }\href
  {\doibase 10.1103/PhysRevLett.80.4510} {\bibfield  {journal} {\bibinfo
  {journal} {Phys. Rev. Lett.}\ }\textbf {\bibinfo {volume} {80}},\ \bibinfo
  {pages} {4510--4513} (\bibinfo {year} {1998})}\BibitemShut {NoStop}%
\bibitem [{\citenamefont {Hedin}(1965)}]{H65}%
  \BibitemOpen
  \bibfield  {author} {\bibinfo {author} {\bibfnamefont {L.}~\bibnamefont
  {Hedin}},\ }\href {\doibase 10.1103/PhysRev.139.A796} {\bibfield  {journal}
  {\bibinfo  {journal} {Phys. Rev.}\ }\textbf {\bibinfo {volume} {139}},\
  \bibinfo {pages} {A796--A823} (\bibinfo {year} {1965})}\BibitemShut {NoStop}%
\bibitem [{\citenamefont {Hybertsen}\ and\ \citenamefont {Louie}(1986)}]{HL86}%
  \BibitemOpen
  \bibfield  {author} {\bibinfo {author} {\bibfnamefont {M.~S.}\ \bibnamefont
  {Hybertsen}}\ and\ \bibinfo {author} {\bibfnamefont {S.~G.}\ \bibnamefont
  {Louie}},\ }\href {\doibase 10.1103/PhysRevB.34.5390} {\bibfield  {journal}
  {\bibinfo  {journal} {Phys. Rev. B}\ }\textbf {\bibinfo {volume} {34}},\
  \bibinfo {pages} {5390--5413} (\bibinfo {year} {1986})}\BibitemShut {NoStop}%
\bibitem [{\citenamefont {Biega}\ \emph {et~al.}(2023)\citenamefont {Biega},
  \citenamefont {Chen}, \citenamefont {Filip},\ and\ \citenamefont
  {Leppert}}]{BCFL23}%
  \BibitemOpen
  \bibfield  {author} {\bibinfo {author} {\bibfnamefont {R.}~\bibnamefont
  {Biega}}, \bibinfo {author} {\bibfnamefont {Y.}~\bibnamefont {Chen}},
  \bibinfo {author} {\bibfnamefont {M.~R.}\ \bibnamefont {Filip}}, \ and\
  \bibinfo {author} {\bibfnamefont {L.}~\bibnamefont {Leppert}},\ }\href
  {\doibase 10.1021/acs.nanolett.3c02285} {\bibfield  {journal} {\bibinfo
  {journal} {Nano Lett.}\ }\textbf {\bibinfo {volume} {23}},\ \bibinfo {pages}
  {8155--8161} (\bibinfo {year} {2023})}\BibitemShut {NoStop}%
\bibitem [{\citenamefont {Bruneval}\ and\ \citenamefont
  {Marques}(2013)}]{BM13}%
  \BibitemOpen
  \bibfield  {author} {\bibinfo {author} {\bibfnamefont {F.}~\bibnamefont
  {Bruneval}}\ and\ \bibinfo {author} {\bibfnamefont {M.~A.~L.}\ \bibnamefont
  {Marques}},\ }\href {\doibase 10.1021/ct300835h} {\bibfield  {journal}
  {\bibinfo  {journal} {J. Chem. Theory Comput.}\ }\textbf {\bibinfo {volume}
  {9}},\ \bibinfo {pages} {324--329} (\bibinfo {year} {2013})}\BibitemShut
  {NoStop}%
\bibitem [{\citenamefont {Fuchs}\ \emph {et~al.}(2007)\citenamefont {Fuchs},
  \citenamefont {Furthm\"uller}, \citenamefont {Bechstedt}, \citenamefont
  {Shishkin},\ and\ \citenamefont {Kresse}}]{FFBS07}%
  \BibitemOpen
  \bibfield  {author} {\bibinfo {author} {\bibfnamefont {F.}~\bibnamefont
  {Fuchs}}, \bibinfo {author} {\bibfnamefont {J.}~\bibnamefont
  {Furthm\"uller}}, \bibinfo {author} {\bibfnamefont {F.}~\bibnamefont
  {Bechstedt}}, \bibinfo {author} {\bibfnamefont {M.}~\bibnamefont {Shishkin}},
  \ and\ \bibinfo {author} {\bibfnamefont {G.}~\bibnamefont {Kresse}},\ }\href
  {\doibase 10.1103/PhysRevB.76.115109} {\bibfield  {journal} {\bibinfo
  {journal} {Phys. Rev. B}\ }\textbf {\bibinfo {volume} {76}},\ \bibinfo
  {pages} {115109} (\bibinfo {year} {2007})}\BibitemShut {NoStop}%
\bibitem [{\citenamefont {Leppert}\ \emph {et~al.}(2019)\citenamefont
  {Leppert}, \citenamefont {Rangel},\ and\ \citenamefont {Neaton}}]{LRN19}%
  \BibitemOpen
  \bibfield  {author} {\bibinfo {author} {\bibfnamefont {L.}~\bibnamefont
  {Leppert}}, \bibinfo {author} {\bibfnamefont {T.}~\bibnamefont {Rangel}}, \
  and\ \bibinfo {author} {\bibfnamefont {J.~B.}\ \bibnamefont {Neaton}},\
  }\href@noop {} {\bibfield  {journal} {\bibinfo  {journal} {Phys. Rev.
  Materials}\ }\textbf {\bibinfo {volume} {3}},\ \bibinfo {pages} {103803}
  (\bibinfo {year} {2019})}\BibitemShut {NoStop}%
\bibitem [{\citenamefont {van Setten}\ \emph {et~al.}(2017)\citenamefont {van
  Setten}, \citenamefont {Giantomassi}, \citenamefont {Gonze}, \citenamefont
  {Rignanese},\ and\ \citenamefont {Hautier}}]{VGGR17}%
  \BibitemOpen
  \bibfield  {author} {\bibinfo {author} {\bibfnamefont {M.~J.}\ \bibnamefont
  {van Setten}}, \bibinfo {author} {\bibfnamefont {M.}~\bibnamefont
  {Giantomassi}}, \bibinfo {author} {\bibfnamefont {X.}~\bibnamefont {Gonze}},
  \bibinfo {author} {\bibfnamefont {G.-M.}\ \bibnamefont {Rignanese}}, \ and\
  \bibinfo {author} {\bibfnamefont {G.}~\bibnamefont {Hautier}},\ }\href
  {\doibase 10.1103/PhysRevB.96.155207} {\bibfield  {journal} {\bibinfo
  {journal} {Phys. Rev. B}\ }\textbf {\bibinfo {volume} {96}},\ \bibinfo
  {pages} {155207} (\bibinfo {year} {2017})}\BibitemShut {NoStop}%
\bibitem [{\citenamefont {Rinke}\ \emph {et~al.}(2005)\citenamefont {Rinke},
  \citenamefont {Qteish}, \citenamefont {Neugebauer}, \citenamefont
  {Freysoldt},\ and\ \citenamefont {Scheffler}}]{RQNF05}%
  \BibitemOpen
  \bibfield  {author} {\bibinfo {author} {\bibfnamefont {P.}~\bibnamefont
  {Rinke}}, \bibinfo {author} {\bibfnamefont {A.}~\bibnamefont {Qteish}},
  \bibinfo {author} {\bibfnamefont {J.}~\bibnamefont {Neugebauer}}, \bibinfo
  {author} {\bibfnamefont {C.}~\bibnamefont {Freysoldt}}, \ and\ \bibinfo
  {author} {\bibfnamefont {M.}~\bibnamefont {Scheffler}},\ }\href {\doibase
  10.1088/1367-2630/7/1/126} {\bibfield  {journal} {\bibinfo  {journal} {NJP}\
  }\textbf {\bibinfo {volume} {7}},\ \bibinfo {pages} {126} (\bibinfo {year}
  {2005})}\BibitemShut {NoStop}%
\bibitem [{\citenamefont {Perdew}\ \emph {et~al.}(2017)\citenamefont {Perdew},
  \citenamefont {Yang}, \citenamefont {Burke}, \citenamefont {Yang},
  \citenamefont {Gross}, \citenamefont {Scheffler}, \citenamefont {Scuseria},
  \citenamefont {Henderson}, \citenamefont {Zhang}, \citenamefont {Ruzsinszky},
  \citenamefont {Peng}, \citenamefont {Sun}, \citenamefont {Trushin},\ and\
  \citenamefont {Görling}}]{PYBY17}%
  \BibitemOpen
  \bibfield  {author} {\bibinfo {author} {\bibfnamefont {J.~P.}\ \bibnamefont
  {Perdew}}, \bibinfo {author} {\bibfnamefont {W.}~\bibnamefont {Yang}},
  \bibinfo {author} {\bibfnamefont {K.}~\bibnamefont {Burke}}, \bibinfo
  {author} {\bibfnamefont {Z.}~\bibnamefont {Yang}}, \bibinfo {author}
  {\bibfnamefont {E.~K.~U.}\ \bibnamefont {Gross}}, \bibinfo {author}
  {\bibfnamefont {M.}~\bibnamefont {Scheffler}}, \bibinfo {author}
  {\bibfnamefont {G.~E.}\ \bibnamefont {Scuseria}}, \bibinfo {author}
  {\bibfnamefont {T.~M.}\ \bibnamefont {Henderson}}, \bibinfo {author}
  {\bibfnamefont {I.~Y.}\ \bibnamefont {Zhang}}, \bibinfo {author}
  {\bibfnamefont {A.}~\bibnamefont {Ruzsinszky}}, \bibinfo {author}
  {\bibfnamefont {H.}~\bibnamefont {Peng}}, \bibinfo {author} {\bibfnamefont
  {J.}~\bibnamefont {Sun}}, \bibinfo {author} {\bibfnamefont {E.}~\bibnamefont
  {Trushin}}, \ and\ \bibinfo {author} {\bibfnamefont {A.}~\bibnamefont
  {Görling}},\ }\href {\doibase 10.1073/pnas.1621352114} {\bibfield  {journal}
  {\bibinfo  {journal} {PNAS}\ }\textbf {\bibinfo {volume} {114}},\ \bibinfo
  {pages} {2801--2806} (\bibinfo {year} {2017})}\BibitemShut {NoStop}%
\bibitem [{\citenamefont {Runge}\ and\ \citenamefont {Gross}(1984)}]{RG84}%
  \BibitemOpen
  \bibfield  {author} {\bibinfo {author} {\bibfnamefont {E.}~\bibnamefont
  {Runge}}\ and\ \bibinfo {author} {\bibfnamefont {E.~K.~U.}\ \bibnamefont
  {Gross}},\ }\href {\doibase 10.1103/PhysRevLett.52.997} {\bibfield  {journal}
  {\bibinfo  {journal} {Phys. Rev. Lett.}\ }\textbf {\bibinfo {volume} {52}},\
  \bibinfo {pages} {997--1000} (\bibinfo {year} {1984})}\BibitemShut {NoStop}%
\bibitem [{\citenamefont {Ullrich}(2012)}]{CU12}%
  \BibitemOpen
  \bibfield  {author} {\bibinfo {author} {\bibfnamefont {C.~A.}\ \bibnamefont
  {Ullrich}},\ }\href@noop {} {\emph {\bibinfo {title} {Time-dependent
  density-functional the- ory: concepts and applications}}}\ (\bibinfo
  {publisher} {Oxford},\ \bibinfo {year} {2012})\BibitemShut {NoStop}%
\bibitem [{\citenamefont {Maitra}(2016)}]{M16}%
  \BibitemOpen
  \bibfield  {author} {\bibinfo {author} {\bibfnamefont {N.~T.}\ \bibnamefont
  {Maitra}},\ }\href {\doibase 10.1063/1.4953039} {\bibfield  {journal}
  {\bibinfo  {journal} {J. Chem. Phys.}\ }\textbf {\bibinfo {volume} {144}},\
  \bibinfo {pages} {220901} (\bibinfo {year} {2016})}\BibitemShut {NoStop}%
\bibitem [{\citenamefont {Perdew}\ \emph {et~al.}(1982)\citenamefont {Perdew},
  \citenamefont {Parr}, \citenamefont {Levy},\ and\ \citenamefont
  {Balduz}}]{PPLB82}%
  \BibitemOpen
  \bibfield  {author} {\bibinfo {author} {\bibfnamefont {J.~P.}\ \bibnamefont
  {Perdew}}, \bibinfo {author} {\bibfnamefont {R.~G.}\ \bibnamefont {Parr}},
  \bibinfo {author} {\bibfnamefont {M.}~\bibnamefont {Levy}}, \ and\ \bibinfo
  {author} {\bibfnamefont {J.~L.}\ \bibnamefont {Balduz}},\ }\href {\doibase
  10.1103/PhysRevLett.49.1691} {\bibfield  {journal} {\bibinfo  {journal}
  {Phys. Rev. Lett.}\ }\textbf {\bibinfo {volume} {49}},\ \bibinfo {pages}
  {1691--1694} (\bibinfo {year} {1982})}\BibitemShut {NoStop}%
\bibitem [{\citenamefont {Stein}\ \emph {et~al.}(2012)\citenamefont {Stein},
  \citenamefont {Autschbach}, \citenamefont {Govind}, \citenamefont {Kronik},\
  and\ \citenamefont {Baer}}]{SAGK12}%
  \BibitemOpen
  \bibfield  {author} {\bibinfo {author} {\bibfnamefont {T.}~\bibnamefont
  {Stein}}, \bibinfo {author} {\bibfnamefont {J.}~\bibnamefont {Autschbach}},
  \bibinfo {author} {\bibfnamefont {N.}~\bibnamefont {Govind}}, \bibinfo
  {author} {\bibfnamefont {L.}~\bibnamefont {Kronik}}, \ and\ \bibinfo {author}
  {\bibfnamefont {R.}~\bibnamefont {Baer}},\ }\href {\doibase
  10.1021/jz3015937} {\bibfield  {journal} {\bibinfo  {journal} {J. Phys. Chem.
  Lett.}\ }\textbf {\bibinfo {volume} {3}},\ \bibinfo {pages} {3740--3744}
  (\bibinfo {year} {2012})}\BibitemShut {NoStop}%
\bibitem [{\citenamefont {Kronik}\ \emph {et~al.}(2012)\citenamefont {Kronik},
  \citenamefont {Stein}, \citenamefont {Refaely-Abramson},\ and\ \citenamefont
  {Baer}}]{KSRB12}%
  \BibitemOpen
  \bibfield  {author} {\bibinfo {author} {\bibfnamefont {L.}~\bibnamefont
  {Kronik}}, \bibinfo {author} {\bibfnamefont {T.}~\bibnamefont {Stein}},
  \bibinfo {author} {\bibfnamefont {S.}~\bibnamefont {Refaely-Abramson}}, \
  and\ \bibinfo {author} {\bibfnamefont {R.}~\bibnamefont {Baer}},\ }\href
  {\doibase 10.1021/ct2009363} {\bibfield  {journal} {\bibinfo  {journal} {J.
  Chem. Theory Comput.}\ }\textbf {\bibinfo {volume} {8}},\ \bibinfo {pages}
  {1515--1531} (\bibinfo {year} {2012})}\BibitemShut {NoStop}%
\bibitem [{\citenamefont {Perdew}\ and\ \citenamefont {Levy}(1983)}]{PL84}%
  \BibitemOpen
  \bibfield  {author} {\bibinfo {author} {\bibfnamefont {J.~P.}\ \bibnamefont
  {Perdew}}\ and\ \bibinfo {author} {\bibfnamefont {M.}~\bibnamefont {Levy}},\
  }\href {\doibase 10.1103/PhysRevLett.51.1884} {\bibfield  {journal} {\bibinfo
   {journal} {Phys. Rev. Lett.}\ }\textbf {\bibinfo {volume} {51}},\ \bibinfo
  {pages} {1884--1887} (\bibinfo {year} {1983})}\BibitemShut {NoStop}%
\bibitem [{\citenamefont {Sham}\ and\ \citenamefont {Schl\"uter}(1983)}]{SS83}%
  \BibitemOpen
  \bibfield  {author} {\bibinfo {author} {\bibfnamefont {L.~J.}\ \bibnamefont
  {Sham}}\ and\ \bibinfo {author} {\bibfnamefont {M.}~\bibnamefont
  {Schl\"uter}},\ }\href {\doibase 10.1103/PhysRevLett.51.1888} {\bibfield
  {journal} {\bibinfo  {journal} {Phys. Rev. Lett.}\ }\textbf {\bibinfo
  {volume} {51}},\ \bibinfo {pages} {1888--1891} (\bibinfo {year}
  {1983})}\BibitemShut {NoStop}%
\bibitem [{\citenamefont {Filip}\ \emph {et~al.}(2016)\citenamefont {Filip},
  \citenamefont {Hillman}, \citenamefont {Haghighirad}, \citenamefont
  {Snaith},\ and\ \citenamefont {Giustino}}]{FHHA16}%
  \BibitemOpen
  \bibfield  {author} {\bibinfo {author} {\bibfnamefont {M.~R.}\ \bibnamefont
  {Filip}}, \bibinfo {author} {\bibfnamefont {S.}~\bibnamefont {Hillman}},
  \bibinfo {author} {\bibfnamefont {A.~A.}\ \bibnamefont {Haghighirad}},
  \bibinfo {author} {\bibfnamefont {H.~J.}\ \bibnamefont {Snaith}}, \ and\
  \bibinfo {author} {\bibfnamefont {F.}~\bibnamefont {Giustino}},\ }\href
  {\doibase 10.1021/acs.jpclett.6b01041} {\bibfield  {journal} {\bibinfo
  {journal} {J. Phys. Chem. Lett.}\ }\textbf {\bibinfo {volume} {7}},\ \bibinfo
  {pages} {2579--2585} (\bibinfo {year} {2016})}\BibitemShut {NoStop}%
\bibitem [{\citenamefont {Slavney}\ \emph
  {et~al.}(2016{\natexlab{b}})\citenamefont {Slavney}, \citenamefont {Hu},
  \citenamefont {Lindenberg},\ and\ \citenamefont {Karunadasa}}]{SHLH16}%
  \BibitemOpen
  \bibfield  {author} {\bibinfo {author} {\bibfnamefont {A.~H.}\ \bibnamefont
  {Slavney}}, \bibinfo {author} {\bibfnamefont {T.}~\bibnamefont {Hu}},
  \bibinfo {author} {\bibfnamefont {A.~M.}\ \bibnamefont {Lindenberg}}, \ and\
  \bibinfo {author} {\bibfnamefont {H.~I.}\ \bibnamefont {Karunadasa}},\ }\href
  {\doibase 10.1021/jacs.5b13294} {\bibfield  {journal} {\bibinfo  {journal}
  {J. Am. Chem. Soc.}\ }\textbf {\bibinfo {volume} {138}},\ \bibinfo {pages}
  {2138--2141} (\bibinfo {year} {2016}{\natexlab{b}})}\BibitemShut {NoStop}%
\bibitem [{\citenamefont {Wright}\ \emph {et~al.}(2021)\citenamefont {Wright},
  \citenamefont {Buizza}, \citenamefont {Savill}, \citenamefont {Longo},
  \citenamefont {Snaith}, \citenamefont {Johnston},\ and\ \citenamefont
  {Herz}}]{WBSL2021}%
  \BibitemOpen
  \bibfield  {author} {\bibinfo {author} {\bibfnamefont {A.~D.}\ \bibnamefont
  {Wright}}, \bibinfo {author} {\bibfnamefont {L.~R.~V.}\ \bibnamefont
  {Buizza}}, \bibinfo {author} {\bibfnamefont {K.~J.}\ \bibnamefont {Savill}},
  \bibinfo {author} {\bibfnamefont {G.}~\bibnamefont {Longo}}, \bibinfo
  {author} {\bibfnamefont {H.~J.}\ \bibnamefont {Snaith}}, \bibinfo {author}
  {\bibfnamefont {M.~B.}\ \bibnamefont {Johnston}}, \ and\ \bibinfo {author}
  {\bibfnamefont {L.~M.}\ \bibnamefont {Herz}},\ }\href {\doibase
  10.1021/acs.jpclett.1c00653} {\bibfield  {journal} {\bibinfo  {journal} {J.
  Phys. Chem. Lett.}\ }\textbf {\bibinfo {volume} {12}},\ \bibinfo {pages}
  {3352--3360} (\bibinfo {year} {2021})}\BibitemShut {NoStop}%
\bibitem [{\citenamefont {Wing}\ \emph {et~al.}(2021)\citenamefont {Wing},
  \citenamefont {Ohad}, \citenamefont {Haber}, \citenamefont {Filip},
  \citenamefont {Gant}, \citenamefont {Neaton},\ and\ \citenamefont
  {Kronik}}]{WOHF21}%
  \BibitemOpen
  \bibfield  {author} {\bibinfo {author} {\bibfnamefont {D.}~\bibnamefont
  {Wing}}, \bibinfo {author} {\bibfnamefont {G.}~\bibnamefont {Ohad}}, \bibinfo
  {author} {\bibfnamefont {J.~B.}\ \bibnamefont {Haber}}, \bibinfo {author}
  {\bibfnamefont {M.~R.}\ \bibnamefont {Filip}}, \bibinfo {author}
  {\bibfnamefont {S.~E.}\ \bibnamefont {Gant}}, \bibinfo {author}
  {\bibfnamefont {J.~B.}\ \bibnamefont {Neaton}}, \ and\ \bibinfo {author}
  {\bibfnamefont {L.}~\bibnamefont {Kronik}},\ }\href {\doibase
  10.1073/pnas.2104556118} {\bibfield  {journal} {\bibinfo  {journal} {PNAS}\
  }\textbf {\bibinfo {volume} {118}},\ \bibinfo {pages} {e2104556118} (\bibinfo
  {year} {2021})}\BibitemShut {NoStop}%
\bibitem [{\citenamefont {Ohad}\ \emph {et~al.}(2022)\citenamefont {Ohad},
  \citenamefont {Wing}, \citenamefont {Gant}, \citenamefont {Cohen},
  \citenamefont {Haber}, \citenamefont {Sagredo}, \citenamefont {Filip},
  \citenamefont {Neaton},\ and\ \citenamefont {Kronik}}]{OWGC22}%
  \BibitemOpen
  \bibfield  {author} {\bibinfo {author} {\bibfnamefont {G.}~\bibnamefont
  {Ohad}}, \bibinfo {author} {\bibfnamefont {D.}~\bibnamefont {Wing}}, \bibinfo
  {author} {\bibfnamefont {S.~E.}\ \bibnamefont {Gant}}, \bibinfo {author}
  {\bibfnamefont {A.~V.}\ \bibnamefont {Cohen}}, \bibinfo {author}
  {\bibfnamefont {J.~B.}\ \bibnamefont {Haber}}, \bibinfo {author}
  {\bibfnamefont {F.}~\bibnamefont {Sagredo}}, \bibinfo {author} {\bibfnamefont
  {M.~R.}\ \bibnamefont {Filip}}, \bibinfo {author} {\bibfnamefont {J.~B.}\
  \bibnamefont {Neaton}}, \ and\ \bibinfo {author} {\bibfnamefont
  {L.}~\bibnamefont {Kronik}},\ }\href {\doibase
  10.1103/PhysRevMaterials.6.104606} {\bibfield  {journal} {\bibinfo  {journal}
  {Phys. Rev. Mater.}\ }\textbf {\bibinfo {volume} {6}},\ \bibinfo {pages}
  {104606} (\bibinfo {year} {2022})}\BibitemShut {NoStop}%
\bibitem [{\citenamefont {Ohad}\ \emph {et~al.}(2023)\citenamefont {Ohad},
  \citenamefont {Gant}, \citenamefont {Wing}, \citenamefont {Haber},
  \citenamefont {{Camarasa-G{\'o}mez}}, \citenamefont {Sagredo}, \citenamefont
  {Filip}, \citenamefont {Neaton},\ and\ \citenamefont {Kronik}}]{OG2023}%
  \BibitemOpen
  \bibfield  {author} {\bibinfo {author} {\bibfnamefont {G.}~\bibnamefont
  {Ohad}}, \bibinfo {author} {\bibfnamefont {S.~E.}\ \bibnamefont {Gant}},
  \bibinfo {author} {\bibfnamefont {D.}~\bibnamefont {Wing}}, \bibinfo {author}
  {\bibfnamefont {Jonah~B.}\ \bibnamefont {Haber}}, \bibinfo {author}
  {\bibfnamefont {Mar{\'i}a}\ \bibnamefont {{Camarasa-G{\'o}mez}}}, \bibinfo
  {author} {\bibfnamefont {F.}~\bibnamefont {Sagredo}}, \bibinfo {author}
  {\bibfnamefont {M.~R.}\ \bibnamefont {Filip}}, \bibinfo {author}
  {\bibfnamefont {J.~B.}\ \bibnamefont {Neaton}}, \ and\ \bibinfo {author}
  {\bibfnamefont {L.}~\bibnamefont {Kronik}},\ }\href {\doibase
  10.1103/PhysRevMaterials.7.123803} {\bibfield  {journal} {\bibinfo  {journal}
  {Phys. Rev. Mater.}\ }\textbf {\bibinfo {volume} {7}},\ \bibinfo {pages}
  {123803} (\bibinfo {year} {2023})}\BibitemShut {NoStop}%
\bibitem [{\citenamefont {Gant}\ \emph {et~al.}(2022)\citenamefont {Gant},
  \citenamefont {Haber}, \citenamefont {Filip}, \citenamefont {Sagredo},
  \citenamefont {Wing}, \citenamefont {Ohad}, \citenamefont {Kronik},\ and\
  \citenamefont {Neaton}}]{GHFS22}%
  \BibitemOpen
  \bibfield  {author} {\bibinfo {author} {\bibfnamefont {S.~E.}\ \bibnamefont
  {Gant}}, \bibinfo {author} {\bibfnamefont {J.~B.}\ \bibnamefont {Haber}},
  \bibinfo {author} {\bibfnamefont {M.~R.}\ \bibnamefont {Filip}}, \bibinfo
  {author} {\bibfnamefont {F.}~\bibnamefont {Sagredo}}, \bibinfo {author}
  {\bibfnamefont {D.}~\bibnamefont {Wing}}, \bibinfo {author} {\bibfnamefont
  {G.}~\bibnamefont {Ohad}}, \bibinfo {author} {\bibfnamefont {L.}~\bibnamefont
  {Kronik}}, \ and\ \bibinfo {author} {\bibfnamefont {J.~B.}\ \bibnamefont
  {Neaton}},\ }\href@noop {} {\bibfield  {journal} {\bibinfo  {journal} {Phys.
  Rev. Materials}\ }\textbf {\bibinfo {volume} {6}},\ \bibinfo {pages} {053802}
  (\bibinfo {year} {2022})}\BibitemShut {NoStop}%
\bibitem [{\citenamefont {Yang}\ and\ \citenamefont {Ullrich}(2013)}]{YU13}%
  \BibitemOpen
  \bibfield  {author} {\bibinfo {author} {\bibfnamefont {Zeng-hui}\
  \bibnamefont {Yang}}\ and\ \bibinfo {author} {\bibfnamefont {Carsten~A.}\
  \bibnamefont {Ullrich}},\ }\bibfield  {title} {\enquote {\bibinfo {title}
  {Direct calculation of exciton binding energies with time-dependent
  density-functional theory},}\ }\href {\doibase 10.1103/PhysRevB.87.195204}
  {\bibfield  {journal} {\bibinfo  {journal} {Phys. Rev. B}\ }\textbf {\bibinfo
  {volume} {87}},\ \bibinfo {pages} {195204} (\bibinfo {year}
  {2013})}\BibitemShut {NoStop}%
\bibitem [{\citenamefont {Zheng}\ \emph {et~al.}(2019)\citenamefont {Zheng},
  \citenamefont {Govoni},\ and\ \citenamefont {Galli}}]{ZGG19}%
  \BibitemOpen
  \bibfield  {author} {\bibinfo {author} {\bibfnamefont {H.}~\bibnamefont
  {Zheng}}, \bibinfo {author} {\bibfnamefont {M.}~\bibnamefont {Govoni}}, \
  and\ \bibinfo {author} {\bibfnamefont {G.}~\bibnamefont {Galli}},\ }\href
  {\doibase 10.1103/PhysRevMaterials.3.073803} {\bibfield  {journal} {\bibinfo
  {journal} {Phys. Rev. Mater.}\ }\textbf {\bibinfo {volume} {3}},\ \bibinfo
  {pages} {073803} (\bibinfo {year} {2019})}\BibitemShut {NoStop}%
\bibitem [{\citenamefont {Chen}\ \emph {et~al.}(2018)\citenamefont {Chen},
  \citenamefont {Miceli}, \citenamefont {Rignanese},\ and\ \citenamefont
  {Pasquarello}}]{CMRP18}%
  \BibitemOpen
  \bibfield  {author} {\bibinfo {author} {\bibfnamefont {W.}~\bibnamefont
  {Chen}}, \bibinfo {author} {\bibfnamefont {G.}~\bibnamefont {Miceli}},
  \bibinfo {author} {\bibfnamefont {G.}~\bibnamefont {Rignanese}}, \ and\
  \bibinfo {author} {\bibfnamefont {A.}~\bibnamefont {Pasquarello}},\ }\href
  {\doibase 10.1103/PhysRevMaterials.2.073803} {\bibfield  {journal} {\bibinfo
  {journal} {Phys. Rev. Mater.}\ }\textbf {\bibinfo {volume} {2}},\ \bibinfo
  {pages} {073803} (\bibinfo {year} {2018})}\BibitemShut {NoStop}%
\bibitem [{\citenamefont {Tal}\ \emph {et~al.}(2020)\citenamefont {Tal},
  \citenamefont {Liu}, \citenamefont {Kresse},\ and\ \citenamefont
  {Pasquarello}}]{TLKP20}%
  \BibitemOpen
  \bibfield  {author} {\bibinfo {author} {\bibfnamefont {A.}~\bibnamefont
  {Tal}}, \bibinfo {author} {\bibfnamefont {P.}~\bibnamefont {Liu}}, \bibinfo
  {author} {\bibfnamefont {G.}~\bibnamefont {Kresse}}, \ and\ \bibinfo {author}
  {\bibfnamefont {A.}~\bibnamefont {Pasquarello}},\ }\href {\doibase
  10.1103/PhysRevResearch.2.032019} {\bibfield  {journal} {\bibinfo  {journal}
  {Phys. Rev. Res.}\ }\textbf {\bibinfo {volume} {2}},\ \bibinfo {pages}
  {032019} (\bibinfo {year} {2020})}\BibitemShut {NoStop}%
\bibitem [{\citenamefont {Dabo}\ \emph {et~al.}(2010)\citenamefont {Dabo},
  \citenamefont {Ferretti}, \citenamefont {Poilvert}, \citenamefont {Li},
  \citenamefont {Marzari},\ and\ \citenamefont {Cococcioni}}]{DFPL10}%
  \BibitemOpen
  \bibfield  {author} {\bibinfo {author} {\bibfnamefont {I.}~\bibnamefont
  {Dabo}}, \bibinfo {author} {\bibfnamefont {A.}~\bibnamefont {Ferretti}},
  \bibinfo {author} {\bibfnamefont {N.}~\bibnamefont {Poilvert}}, \bibinfo
  {author} {\bibfnamefont {Y.}~\bibnamefont {Li}}, \bibinfo {author}
  {\bibfnamefont {N.}~\bibnamefont {Marzari}}, \ and\ \bibinfo {author}
  {\bibfnamefont {M.}~\bibnamefont {Cococcioni}},\ }\href {\doibase
  10.1103/PhysRevB.82.115121} {\bibfield  {journal} {\bibinfo  {journal} {Phys.
  Rev. B}\ }\textbf {\bibinfo {volume} {82}},\ \bibinfo {pages} {115121}
  (\bibinfo {year} {2010})}\BibitemShut {NoStop}%
\bibitem [{\citenamefont {Dabo}\ \emph {et~al.}(2014)\citenamefont {Dabo},
  \citenamefont {Ferretti},\ and\ \citenamefont {Marzari}}]{DFM14}%
  \BibitemOpen
  \bibfield  {author} {\bibinfo {author} {\bibfnamefont {I.}~\bibnamefont
  {Dabo}}, \bibinfo {author} {\bibfnamefont {A.}~\bibnamefont {Ferretti}}, \
  and\ \bibinfo {author} {\bibfnamefont {N.}~\bibnamefont {Marzari}},\
  }\enquote {\bibinfo {title} {Piecewise linearity and spectroscopic properties
  from koopmans-compliant functionals},}\ in\ \href {\doibase
  10.1007/128_2013_504} {\emph {\bibinfo {booktitle} {First Principles
  Approaches to Spectroscopic Properties of Complex Materials}}}\ (\bibinfo
  {publisher} {Springer Berlin Heidelberg},\ \bibinfo {address} {Berlin,
  Heidelberg},\ \bibinfo {year} {2014})\ pp.\ \bibinfo {pages}
  {193--233}\BibitemShut {NoStop}%
\bibitem [{\citenamefont {Miceli}\ \emph {et~al.}(2018)\citenamefont {Miceli},
  \citenamefont {Chen}, \citenamefont {Reshetnyak},\ and\ \citenamefont
  {Pasquarello}}]{MCRP18}%
  \BibitemOpen
  \bibfield  {author} {\bibinfo {author} {\bibfnamefont {G.}~\bibnamefont
  {Miceli}}, \bibinfo {author} {\bibfnamefont {W.}~\bibnamefont {Chen}},
  \bibinfo {author} {\bibfnamefont {I.}~\bibnamefont {Reshetnyak}}, \ and\
  \bibinfo {author} {\bibfnamefont {A.}~\bibnamefont {Pasquarello}},\ }\href
  {\doibase 10.1103/PhysRevB.97.121112} {\bibfield  {journal} {\bibinfo
  {journal} {Phys. Rev. B}\ }\textbf {\bibinfo {volume} {97}},\ \bibinfo
  {pages} {121112} (\bibinfo {year} {2018})}\BibitemShut {NoStop}%
\bibitem [{\citenamefont {Nguyen}\ \emph {et~al.}(2018)\citenamefont {Nguyen},
  \citenamefont {Colonna}, \citenamefont {Ferretti},\ and\ \citenamefont
  {Marzari}}]{NCFM18}%
  \BibitemOpen
  \bibfield  {author} {\bibinfo {author} {\bibfnamefont {N.~L.}\ \bibnamefont
  {Nguyen}}, \bibinfo {author} {\bibfnamefont {N.}~\bibnamefont {Colonna}},
  \bibinfo {author} {\bibfnamefont {A.}~\bibnamefont {Ferretti}}, \ and\
  \bibinfo {author} {\bibfnamefont {N.}~\bibnamefont {Marzari}},\ }\href
  {\doibase 10.1103/PhysRevX.8.021051} {\bibfield  {journal} {\bibinfo
  {journal} {Phys. Rev. X}\ }\textbf {\bibinfo {volume} {8}},\ \bibinfo {pages}
  {021051} (\bibinfo {year} {2018})}\BibitemShut {NoStop}%
\bibitem [{\citenamefont {Mahler}\ \emph {et~al.}(2022)\citenamefont {Mahler},
  \citenamefont {Williams}, \citenamefont {Su},\ and\ \citenamefont
  {Yang}}]{MWSY22}%
  \BibitemOpen
  \bibfield  {author} {\bibinfo {author} {\bibfnamefont {A.}~\bibnamefont
  {Mahler}}, \bibinfo {author} {\bibfnamefont {J.}~\bibnamefont {Williams}},
  \bibinfo {author} {\bibfnamefont {N.~Q.}\ \bibnamefont {Su}}, \ and\ \bibinfo
  {author} {\bibfnamefont {W.}~\bibnamefont {Yang}},\ }\href {\doibase
  10.1103/PhysRevB.106.035147} {\bibfield  {journal} {\bibinfo  {journal}
  {Phys. Rev. B}\ }\textbf {\bibinfo {volume} {106}},\ \bibinfo {pages}
  {035147} (\bibinfo {year} {2022})}\BibitemShut {NoStop}%
\bibitem [{\citenamefont {Linscott}\ \emph {et~al.}(2023)\citenamefont
  {Linscott}, \citenamefont {Colonna}, \citenamefont {De~Gennaro},
  \citenamefont {Nguyen}, \citenamefont {Borghi}, \citenamefont {Ferretti},
  \citenamefont {Dabo},\ and\ \citenamefont {Marzari}}]{LCDN23}%
  \BibitemOpen
  \bibfield  {author} {\bibinfo {author} {\bibfnamefont {E.~B.}\ \bibnamefont
  {Linscott}}, \bibinfo {author} {\bibfnamefont {N.}~\bibnamefont {Colonna}},
  \bibinfo {author} {\bibfnamefont {R.}~\bibnamefont {De~Gennaro}}, \bibinfo
  {author} {\bibfnamefont {N.~L.}\ \bibnamefont {Nguyen}}, \bibinfo {author}
  {\bibfnamefont {G.}~\bibnamefont {Borghi}}, \bibinfo {author} {\bibfnamefont
  {A.}~\bibnamefont {Ferretti}}, \bibinfo {author} {\bibfnamefont
  {I.}~\bibnamefont {Dabo}}, \ and\ \bibinfo {author} {\bibfnamefont
  {N.}~\bibnamefont {Marzari}},\ }\href {\doibase 10.1021/acs.jctc.3c00652}
  {\bibfield  {journal} {\bibinfo  {journal} {J. Chem. Theory Comput.}\
  }\textbf {\bibinfo {volume} {19}},\ \bibinfo {pages} {7097--7111} (\bibinfo
  {year} {2023})},\ \bibinfo {note} {pMID: 37610300}\BibitemShut {NoStop}%
\bibitem [{\citenamefont {Refaely-Abramson}\ \emph {et~al.}(2013)\citenamefont
  {Refaely-Abramson}, \citenamefont {Sharifzadeh}, \citenamefont {Jain},
  \citenamefont {Baer}, \citenamefont {Neaton},\ and\ \citenamefont
  {Kronik}}]{RSJB13}%
  \BibitemOpen
  \bibfield  {author} {\bibinfo {author} {\bibfnamefont {S.}~\bibnamefont
  {Refaely-Abramson}}, \bibinfo {author} {\bibfnamefont {S.}~\bibnamefont
  {Sharifzadeh}}, \bibinfo {author} {\bibfnamefont {M.}~\bibnamefont {Jain}},
  \bibinfo {author} {\bibfnamefont {R.}~\bibnamefont {Baer}}, \bibinfo {author}
  {\bibfnamefont {J.~B.}\ \bibnamefont {Neaton}}, \ and\ \bibinfo {author}
  {\bibfnamefont {L.}~\bibnamefont {Kronik}},\ }\href {\doibase
  10.1103/PhysRevB.88.081204} {\bibfield  {journal} {\bibinfo  {journal} {Phys.
  Rev. B}\ }\textbf {\bibinfo {volume} {88}},\ \bibinfo {pages} {081204}
  (\bibinfo {year} {2013})}\BibitemShut {NoStop}%
\bibitem [{\citenamefont {Stein}\ \emph {et~al.}(2010)\citenamefont {Stein},
  \citenamefont {Eisenberg}, \citenamefont {Kronik},\ and\ \citenamefont
  {Baer}}]{SEKB10}%
  \BibitemOpen
  \bibfield  {author} {\bibinfo {author} {\bibfnamefont {T.}~\bibnamefont
  {Stein}}, \bibinfo {author} {\bibfnamefont {H.}~\bibnamefont {Eisenberg}},
  \bibinfo {author} {\bibfnamefont {L.}~\bibnamefont {Kronik}}, \ and\ \bibinfo
  {author} {\bibfnamefont {R.}~\bibnamefont {Baer}},\ }\href {\doibase
  10.1103/PhysRevLett.105.266802} {\bibfield  {journal} {\bibinfo  {journal}
  {Phys. Rev. Lett.}\ }\textbf {\bibinfo {volume} {105}},\ \bibinfo {pages}
  {266802} (\bibinfo {year} {2010})}\BibitemShut {NoStop}%
\bibitem [{\citenamefont {Refaely-Abramson}\ \emph {et~al.}(2011)\citenamefont
  {Refaely-Abramson}, \citenamefont {Baer},\ and\ \citenamefont
  {Kronik}}]{RBK11}%
  \BibitemOpen
  \bibfield  {author} {\bibinfo {author} {\bibfnamefont {S.}~\bibnamefont
  {Refaely-Abramson}}, \bibinfo {author} {\bibfnamefont {R.}~\bibnamefont
  {Baer}}, \ and\ \bibinfo {author} {\bibfnamefont {L.}~\bibnamefont
  {Kronik}},\ }\href {\doibase 10.1103/PhysRevB.84.075144} {\bibfield
  {journal} {\bibinfo  {journal} {Phys. Rev. B}\ }\textbf {\bibinfo {volume}
  {84}},\ \bibinfo {pages} {075144} (\bibinfo {year} {2011})}\BibitemShut
  {NoStop}%
\bibitem [{\citenamefont {Autschbach}\ and\ \citenamefont
  {Srebro}(2014)}]{AS14}%
  \BibitemOpen
  \bibfield  {author} {\bibinfo {author} {\bibfnamefont {J.}~\bibnamefont
  {Autschbach}}\ and\ \bibinfo {author} {\bibfnamefont {M.}~\bibnamefont
  {Srebro}},\ }\href {\doibase 10.1021/ar500171t} {\bibfield  {journal}
  {\bibinfo  {journal} {Acc. Chem. Res.}\ }\textbf {\bibinfo {volume} {47}},\
  \bibinfo {pages} {2592--2602} (\bibinfo {year} {2014})}\BibitemShut {NoStop}%
\bibitem [{\citenamefont {Mori-S\'anchez}\ \emph {et~al.}(2008)\citenamefont
  {Mori-S\'anchez}, \citenamefont {Cohen},\ and\ \citenamefont {Yang}}]{MCY08}%
  \BibitemOpen
  \bibfield  {author} {\bibinfo {author} {\bibfnamefont {P.}~\bibnamefont
  {Mori-S\'anchez}}, \bibinfo {author} {\bibfnamefont {A.~J.}\ \bibnamefont
  {Cohen}}, \ and\ \bibinfo {author} {\bibfnamefont {W.}~\bibnamefont {Yang}},\
  }\href {\doibase 10.1103/PhysRevLett.100.146401} {\bibfield  {journal}
  {\bibinfo  {journal} {Phys. Rev. Lett.}\ }\textbf {\bibinfo {volume} {100}},\
  \bibinfo {pages} {146401} (\bibinfo {year} {2008})}\BibitemShut {NoStop}%
\bibitem [{\citenamefont {Kraisler}\ and\ \citenamefont {Kronik}(2014)}]{KK15}%
  \BibitemOpen
  \bibfield  {author} {\bibinfo {author} {\bibfnamefont {E.}~\bibnamefont
  {Kraisler}}\ and\ \bibinfo {author} {\bibfnamefont {L.}~\bibnamefont
  {Kronik}},\ }\href {\doibase 10.1063/1.4871462} {\bibfield  {journal}
  {\bibinfo  {journal} {J. Chem. Phys.}\ }\textbf {\bibinfo {volume} {140}},\
  \bibinfo {pages} {18A540} (\bibinfo {year} {2014})}\BibitemShut {NoStop}%
\bibitem [{\citenamefont {Vlček}\ \emph {et~al.}(2015)\citenamefont {Vlček},
  \citenamefont {Eisenberg}, \citenamefont {Steinle-Neumann}, \citenamefont
  {Kronik},\ and\ \citenamefont {Baer}}]{VESN16}%
  \BibitemOpen
  \bibfield  {author} {\bibinfo {author} {\bibfnamefont {V.}~\bibnamefont
  {Vlček}}, \bibinfo {author} {\bibfnamefont {H.~R.}\ \bibnamefont
  {Eisenberg}}, \bibinfo {author} {\bibfnamefont {G.}~\bibnamefont
  {Steinle-Neumann}}, \bibinfo {author} {\bibfnamefont {L.}~\bibnamefont
  {Kronik}}, \ and\ \bibinfo {author} {\bibfnamefont {R.}~\bibnamefont
  {Baer}},\ }\href {\doibase 10.1063/1.4905236} {\bibfield  {journal} {\bibinfo
   {journal} {J. Chem. Phys.}\ }\textbf {\bibinfo {volume} {142}},\ \bibinfo
  {pages} {034107} (\bibinfo {year} {2015})}\BibitemShut {NoStop}%
\bibitem [{\citenamefont {G\"orling}(2015)}]{G15}%
  \BibitemOpen
  \bibfield  {author} {\bibinfo {author} {\bibfnamefont {A.}~\bibnamefont
  {G\"orling}},\ }\href {\doibase 10.1103/PhysRevB.91.245120} {\bibfield
  {journal} {\bibinfo  {journal} {Phys. Rev. B}\ }\textbf {\bibinfo {volume}
  {91}},\ \bibinfo {pages} {245120} (\bibinfo {year} {2015})}\BibitemShut
  {NoStop}%
\bibitem [{\citenamefont {Ma}\ and\ \citenamefont {Wang}(2016)}]{MW16}%
  \BibitemOpen
  \bibfield  {author} {\bibinfo {author} {\bibfnamefont {J.}~\bibnamefont
  {Ma}}\ and\ \bibinfo {author} {\bibfnamefont {L.}~\bibnamefont {Wang}},\
  }\href {\doibase 10.1038/srep24924} {\bibfield  {journal} {\bibinfo
  {journal} {Sci. Rep.}\ }\textbf {\bibinfo {volume} {6}},\ \bibinfo {pages}
  {24924} (\bibinfo {year} {2016})}\BibitemShut {NoStop}%
\bibitem [{\citenamefont {Makov}\ and\ \citenamefont {Payne}(1995)}]{MP95}%
  \BibitemOpen
  \bibfield  {author} {\bibinfo {author} {\bibfnamefont {G.}~\bibnamefont
  {Makov}}\ and\ \bibinfo {author} {\bibfnamefont {M.~C.}\ \bibnamefont
  {Payne}},\ }\href {\doibase 10.1103/PhysRevB.51.4014} {\bibfield  {journal}
  {\bibinfo  {journal} {Phys. Rev. B}\ }\textbf {\bibinfo {volume} {51}},\
  \bibinfo {pages} {4014--4022} (\bibinfo {year} {1995})}\BibitemShut {NoStop}%
\bibitem [{\citenamefont {Wei}\ \emph {et~al.}(2019)\citenamefont {Wei},
  \citenamefont {Deng}, \citenamefont {Sun}, \citenamefont {Hartono},
  \citenamefont {Seng}, \citenamefont {Buonassisi}, \citenamefont {Bristowe},\
  and\ \citenamefont {Cheetham}}]{WDSH19}%
  \BibitemOpen
  \bibfield  {author} {\bibinfo {author} {\bibfnamefont {F.}~\bibnamefont
  {Wei}}, \bibinfo {author} {\bibfnamefont {Z.}~\bibnamefont {Deng}}, \bibinfo
  {author} {\bibfnamefont {S.}~\bibnamefont {Sun}}, \bibinfo {author}
  {\bibfnamefont {N.~T.~P.}\ \bibnamefont {Hartono}}, \bibinfo {author}
  {\bibfnamefont {H.~L.}\ \bibnamefont {Seng}}, \bibinfo {author}
  {\bibfnamefont {T.}~\bibnamefont {Buonassisi}}, \bibinfo {author}
  {\bibfnamefont {P.~D.}\ \bibnamefont {Bristowe}}, \ and\ \bibinfo {author}
  {\bibfnamefont {A.~K.}\ \bibnamefont {Cheetham}},\ }\href {\doibase
  10.1039/C9CC01134J} {\bibfield  {journal} {\bibinfo  {journal} {Chem.
  Commun.}\ }\textbf {\bibinfo {volume} {55}},\ \bibinfo {pages} {3721--3724}
  (\bibinfo {year} {2019})}\BibitemShut {NoStop}%
\bibitem [{\citenamefont {Tran}\ \emph {et~al.}(2017)\citenamefont {Tran},
  \citenamefont {Panella}, \citenamefont {Chamorro}, \citenamefont {Morey},\
  and\ \citenamefont {McQueen}}]{TPCM17}%
  \BibitemOpen
  \bibfield  {author} {\bibinfo {author} {\bibfnamefont {T.~Thao}\ \bibnamefont
  {Tran}}, \bibinfo {author} {\bibfnamefont {J.~R.}\ \bibnamefont {Panella}},
  \bibinfo {author} {\bibfnamefont {J.~R.}\ \bibnamefont {Chamorro}}, \bibinfo
  {author} {\bibfnamefont {J.~R.}\ \bibnamefont {Morey}}, \ and\ \bibinfo
  {author} {\bibfnamefont {T.~M.}\ \bibnamefont {McQueen}},\ }\href {\doibase
  10.1039/C7MH00239D} {\bibfield  {journal} {\bibinfo  {journal} {Mater.
  Horiz.}\ }\textbf {\bibinfo {volume} {4}},\ \bibinfo {pages} {688--693}
  (\bibinfo {year} {2017})}\BibitemShut {NoStop}%
\bibitem [{\citenamefont {Hybertsen}\ and\ \citenamefont {Louie}(1985)}]{HL85}%
  \BibitemOpen
  \bibfield  {author} {\bibinfo {author} {\bibfnamefont {M.~S.}\ \bibnamefont
  {Hybertsen}}\ and\ \bibinfo {author} {\bibfnamefont {S.~G.}\ \bibnamefont
  {Louie}},\ }\href@noop {} {\bibfield  {journal} {\bibinfo  {journal} {Phys.
  Rev. Lett.}\ }\textbf {\bibinfo {volume} {55}},\ \bibinfo {pages}
  {1418--1421} (\bibinfo {year} {1985})}\BibitemShut {NoStop}%
\bibitem [{\citenamefont {Deslippe}\ \emph {et~al.}(2012)\citenamefont
  {Deslippe}, \citenamefont {Samsonidze}, \citenamefont {Strubbe},
  \citenamefont {Jain}, \citenamefont {Cohen},\ and\ \citenamefont
  {Louie}}]{DSSJ12}%
  \BibitemOpen
  \bibfield  {author} {\bibinfo {author} {\bibfnamefont {J.}~\bibnamefont
  {Deslippe}}, \bibinfo {author} {\bibfnamefont {G.}~\bibnamefont
  {Samsonidze}}, \bibinfo {author} {\bibfnamefont {D.~A.}\ \bibnamefont
  {Strubbe}}, \bibinfo {author} {\bibfnamefont {M.}~\bibnamefont {Jain}},
  \bibinfo {author} {\bibfnamefont {M.~L.}\ \bibnamefont {Cohen}}, \ and\
  \bibinfo {author} {\bibfnamefont {S.~G.}\ \bibnamefont {Louie}},\ }\href
  {\doibase 10.1016/j.cpc.2011.12.006} {\bibfield  {journal} {\bibinfo
  {journal} {Comput. Phys. Commun.}\ }\textbf {\bibinfo {volume} {183}},\
  \bibinfo {pages} {1269--1289} (\bibinfo {year} {2012})}\BibitemShut {NoStop}%
\bibitem [{sup()}]{supplementaryinformation}%
  \BibitemOpen
  \href@noop {} {\enquote {\bibinfo {title} {{See Supplemental Information at
  [URL will be inserted by publisher] for more technical details on
  calculations}},}\ }\BibitemShut {NoStop}%
\bibitem [{\citenamefont {Sancho-Garc{\'i}a}\ \emph {et~al.}(2003)\citenamefont
  {Sancho-Garc{\'i}a}, \citenamefont {Br{\'e}das},\ and\ \citenamefont
  {Cornil}}]{SBC03}%
  \BibitemOpen
  \bibfield  {author} {\bibinfo {author} {\bibfnamefont {J.C}\ \bibnamefont
  {Sancho-Garc{\'i}a}}, \bibinfo {author} {\bibfnamefont {J.L}\ \bibnamefont
  {Br{\'e}das}}, \ and\ \bibinfo {author} {\bibfnamefont {J.}~\bibnamefont
  {Cornil}},\ }\href {\doibase https://doi.org/10.1016/S0009-2614(03)01086-8}
  {\bibfield  {journal} {\bibinfo  {journal} {Chem. Phys. Lett.}\ }\textbf
  {\bibinfo {volume} {377}},\ \bibinfo {pages} {63--68} (\bibinfo {year}
  {2003})}\BibitemShut {NoStop}%
\bibitem [{\citenamefont {Gavartin}\ \emph {et~al.}(2003)\citenamefont
  {Gavartin}, \citenamefont {Sushko},\ and\ \citenamefont {Shluger}}]{GSS03}%
  \BibitemOpen
  \bibfield  {author} {\bibinfo {author} {\bibfnamefont {J.~L.}\ \bibnamefont
  {Gavartin}}, \bibinfo {author} {\bibfnamefont {P.~V.}\ \bibnamefont
  {Sushko}}, \ and\ \bibinfo {author} {\bibfnamefont {A.~L.}\ \bibnamefont
  {Shluger}},\ }\href {\doibase 10.1103/PhysRevB.67.035108} {\bibfield
  {journal} {\bibinfo  {journal} {Phys. Rev. B}\ }\textbf {\bibinfo {volume}
  {67}},\ \bibinfo {pages} {035108} (\bibinfo {year} {2003})}\BibitemShut
  {NoStop}%
\bibitem [{\citenamefont {Lynch}\ \emph {et~al.}(2000)\citenamefont {Lynch},
  \citenamefont {Fast}, \citenamefont {Harris},\ and\ \citenamefont
  {Truhlar}}]{LFHL00}%
  \BibitemOpen
  \bibfield  {author} {\bibinfo {author} {\bibfnamefont {B.~J.}\ \bibnamefont
  {Lynch}}, \bibinfo {author} {\bibfnamefont {P.~L.}\ \bibnamefont {Fast}},
  \bibinfo {author} {\bibfnamefont {M.}~\bibnamefont {Harris}}, \ and\ \bibinfo
  {author} {\bibfnamefont {D.~G.}\ \bibnamefont {Truhlar}},\ }\href {\doibase
  10.1021/jp000497z} {\bibfield  {journal} {\bibinfo  {journal} {J. Phys. Chem.
  A}\ }\textbf {\bibinfo {volume} {104}},\ \bibinfo {pages} {4811--4815}
  (\bibinfo {year} {2000})}\BibitemShut {NoStop}%
\bibitem [{\citenamefont {Varley}\ \emph {et~al.}(2010)\citenamefont {Varley},
  \citenamefont {Janotti},\ and\ \citenamefont {Van~de Walle}}]{VJV10}%
  \BibitemOpen
  \bibfield  {author} {\bibinfo {author} {\bibfnamefont {J.~B.}\ \bibnamefont
  {Varley}}, \bibinfo {author} {\bibfnamefont {A.}~\bibnamefont {Janotti}}, \
  and\ \bibinfo {author} {\bibfnamefont {C.~G.}\ \bibnamefont {Van~de Walle}},\
  }\href {\doibase 10.1103/PhysRevB.81.245216} {\bibfield  {journal} {\bibinfo
  {journal} {Phys. Rev. B}\ }\textbf {\bibinfo {volume} {81}},\ \bibinfo
  {pages} {245216} (\bibinfo {year} {2010})}\BibitemShut {NoStop}%
\bibitem [{\citenamefont {Garrick}\ \emph {et~al.}(2020)\citenamefont
  {Garrick}, \citenamefont {Natan}, \citenamefont {Gould},\ and\ \citenamefont
  {Kronik}}]{GNGK20}%
  \BibitemOpen
  \bibfield  {author} {\bibinfo {author} {\bibfnamefont {R.}~\bibnamefont
  {Garrick}}, \bibinfo {author} {\bibfnamefont {A.}~\bibnamefont {Natan}},
  \bibinfo {author} {\bibfnamefont {T.}~\bibnamefont {Gould}}, \ and\ \bibinfo
  {author} {\bibfnamefont {L.}~\bibnamefont {Kronik}},\ }\href {\doibase
  10.1103/PhysRevX.10.021040} {\bibfield  {journal} {\bibinfo  {journal} {Phys.
  Rev. X}\ }\textbf {\bibinfo {volume} {10}},\ \bibinfo {pages} {021040}
  (\bibinfo {year} {2020})}\BibitemShut {NoStop}%
\bibitem [{\citenamefont {Perdew}\ \emph
  {et~al.}(1996{\natexlab{a}})\citenamefont {Perdew}, \citenamefont {Burke},\
  and\ \citenamefont {Ernzerhof}}]{PBE96}%
  \BibitemOpen
  \bibfield  {author} {\bibinfo {author} {\bibfnamefont {J.~P.}\ \bibnamefont
  {Perdew}}, \bibinfo {author} {\bibfnamefont {K.}~\bibnamefont {Burke}}, \
  and\ \bibinfo {author} {\bibfnamefont {M.}~\bibnamefont {Ernzerhof}},\ }\href
  {https://link.aps.org/doi/10.1103/PhysRevLett.77.3865} {\bibfield  {journal}
  {\bibinfo  {journal} {Phys. Rev. Lett.}\ }\textbf {\bibinfo {volume} {77}},\
  \bibinfo {pages} {3865--3868} (\bibinfo {year}
  {1996}{\natexlab{a}})}\BibitemShut {NoStop}%
\bibitem [{\citenamefont {Perdew}\ \emph
  {et~al.}(1996{\natexlab{b}})\citenamefont {Perdew}, \citenamefont
  {Ernzerhof},\ and\ \citenamefont {Burke}}]{PBE0}%
  \BibitemOpen
  \bibfield  {author} {\bibinfo {author} {\bibfnamefont {J.~P.}\ \bibnamefont
  {Perdew}}, \bibinfo {author} {\bibfnamefont {M.}~\bibnamefont {Ernzerhof}}, \
  and\ \bibinfo {author} {\bibfnamefont {K.}~\bibnamefont {Burke}},\ }\href
  {\doibase 10.1063/1.472933} {\bibfield  {journal} {\bibinfo  {journal} {J.
  Chem. Phys.}\ }\textbf {\bibinfo {volume} {105}},\ \bibinfo {pages}
  {9982--9985} (\bibinfo {year} {1996}{\natexlab{b}})}\BibitemShut {NoStop}%
\bibitem [{\citenamefont {Krukau}\ \emph {et~al.}(2006)\citenamefont {Krukau},
  \citenamefont {Vydrov}, \citenamefont {Izmaylov},\ and\ \citenamefont
  {Scuseria}}]{HSE06}%
  \BibitemOpen
  \bibfield  {author} {\bibinfo {author} {\bibfnamefont {A.~V.}\ \bibnamefont
  {Krukau}}, \bibinfo {author} {\bibfnamefont {O.~A.}\ \bibnamefont {Vydrov}},
  \bibinfo {author} {\bibfnamefont {A.~F.}\ \bibnamefont {Izmaylov}}, \ and\
  \bibinfo {author} {\bibfnamefont {G.~E.}\ \bibnamefont {Scuseria}},\ }\href
  {\doibase 10.1063/1.2404663} {\bibfield  {journal} {\bibinfo  {journal} {J.
  Chem. Phys.}\ }\textbf {\bibinfo {volume} {125}},\ \bibinfo {pages} {224106}
  (\bibinfo {year} {2006})}\BibitemShut {NoStop}%
\bibitem [{\citenamefont {Lebeda}\ \emph {et~al.}(2023)\citenamefont {Lebeda},
  \citenamefont {Aschebrock}, \citenamefont {Sun}, \citenamefont {Leppert},\
  and\ \citenamefont {K\"ummel}}]{LASLK23}%
  \BibitemOpen
  \bibfield  {author} {\bibinfo {author} {\bibfnamefont {Timo}\ \bibnamefont
  {Lebeda}}, \bibinfo {author} {\bibfnamefont {Thilo}\ \bibnamefont
  {Aschebrock}}, \bibinfo {author} {\bibfnamefont {Jianwei}\ \bibnamefont
  {Sun}}, \bibinfo {author} {\bibfnamefont {Linn}\ \bibnamefont {Leppert}}, \
  and\ \bibinfo {author} {\bibfnamefont {Stephan}\ \bibnamefont {K\"ummel}},\
  }\href {\doibase 10.1103/PhysRevMaterials.7.093803} {\bibfield  {journal}
  {\bibinfo  {journal} {Phys. Rev. Mater.}\ }\textbf {\bibinfo {volume} {7}},\
  \bibinfo {pages} {093803} (\bibinfo {year} {2023})}\BibitemShut {NoStop}%
\bibitem [{\citenamefont {Delor}\ \emph {et~al.}(2020)\citenamefont {Delor},
  \citenamefont {Slavney}, \citenamefont {Wolf}, \citenamefont {Filip},
  \citenamefont {Neaton}, \citenamefont {Karunadasa},\ and\ \citenamefont
  {Ginsberg}}]{DSWF20}%
  \BibitemOpen
  \bibfield  {author} {\bibinfo {author} {\bibfnamefont {M.}~\bibnamefont
  {Delor}}, \bibinfo {author} {\bibfnamefont {A.~H.}\ \bibnamefont {Slavney}},
  \bibinfo {author} {\bibfnamefont {N.~R.}\ \bibnamefont {Wolf}}, \bibinfo
  {author} {\bibfnamefont {M.~R.}\ \bibnamefont {Filip}}, \bibinfo {author}
  {\bibfnamefont {J.~B.}\ \bibnamefont {Neaton}}, \bibinfo {author}
  {\bibfnamefont {H.~I.}\ \bibnamefont {Karunadasa}}, \ and\ \bibinfo {author}
  {\bibfnamefont {N.~S.}\ \bibnamefont {Ginsberg}},\ }\href {\doibase
  10.1021/acsenergylett.0c00414} {\bibfield  {journal} {\bibinfo  {journal}
  {ACS Energy Lett.}\ }\textbf {\bibinfo {volume} {5}},\ \bibinfo {pages}
  {1337--1345} (\bibinfo {year} {2020})}\BibitemShut {NoStop}%
\bibitem [{\citenamefont {Elliott}(1957)}]{E57}%
  \BibitemOpen
  \bibfield  {author} {\bibinfo {author} {\bibfnamefont {R.~J.}\ \bibnamefont
  {Elliott}},\ }\href {\doibase 10.1103/PhysRev.108.1384} {\bibfield  {journal}
  {\bibinfo  {journal} {Phys. Rev.}\ }\textbf {\bibinfo {volume} {108}},\
  \bibinfo {pages} {1384--1389} (\bibinfo {year} {1957})}\BibitemShut {NoStop}%
\bibitem [{\citenamefont {Wannier}(1937)}]{W37}%
  \BibitemOpen
  \bibfield  {author} {\bibinfo {author} {\bibfnamefont {G.~H.}\ \bibnamefont
  {Wannier}},\ }\href {\doibase 10.1103/PhysRev.52.191} {\bibfield  {journal}
  {\bibinfo  {journal} {Phys. Rev.}\ }\textbf {\bibinfo {volume} {52}},\
  \bibinfo {pages} {191--197} (\bibinfo {year} {1937})}\BibitemShut {NoStop}%
\bibitem [{\citenamefont {Longo}\ \emph {et~al.}(2020)\citenamefont {Longo},
  \citenamefont {Mahesh}, \citenamefont {Buizza}, \citenamefont {Wright},
  \citenamefont {Ramadan}, \citenamefont {Abdi-Jalebi}, \citenamefont {Nayak},
  \citenamefont {Herz},\ and\ \citenamefont {Snaith}}]{LMBW20}%
  \BibitemOpen
  \bibfield  {author} {\bibinfo {author} {\bibfnamefont {G.}~\bibnamefont
  {Longo}}, \bibinfo {author} {\bibfnamefont {S.}~\bibnamefont {Mahesh}},
  \bibinfo {author} {\bibfnamefont {L.~R.~V.}\ \bibnamefont {Buizza}}, \bibinfo
  {author} {\bibfnamefont {A.~D.}\ \bibnamefont {Wright}}, \bibinfo {author}
  {\bibfnamefont {A.~J.}\ \bibnamefont {Ramadan}}, \bibinfo {author}
  {\bibfnamefont {M.}~\bibnamefont {Abdi-Jalebi}}, \bibinfo {author}
  {\bibfnamefont {P.~K.}\ \bibnamefont {Nayak}}, \bibinfo {author}
  {\bibfnamefont {L.~M.}\ \bibnamefont {Herz}}, \ and\ \bibinfo {author}
  {\bibfnamefont {H.~J.}\ \bibnamefont {Snaith}},\ }\href {\doibase
  10.1021/acsenergylett.0c01020} {\bibfield  {journal} {\bibinfo  {journal}
  {ACS Energy Lett.}\ }\textbf {\bibinfo {volume} {5}},\ \bibinfo {pages}
  {2200--2207} (\bibinfo {year} {2020})}\BibitemShut {NoStop}%
\end{thebibliography}%
\label{page:end}
\end{document}